\documentclass[a4paper,fleqn]{cas-dc}

\usepackage[authoryear,longnamesfirst]{natbib}
\usepackage{adjustbox}
\usepackage[graphicx]{realboxes}
\usepackage{rotating}

\newcommand{\isis}{\textit{ISIS} }
\newcommand{\czerny}{Schwarzenberg-Czerny }
\newcommand{\degs}{deg$^2$ }

\def\tsc#1{\csdef{#1}{\textsc{\lowercase{#1}}\xspace}}
\tsc{WGM}
\tsc{QE}

\begin{document}
\let\WriteBookmarks\relax
\def\floatpagepagefraction{1}
\def\textpagefraction{.001}

\shorttitle{New variable sources in LMC}    

\shortauthors{A. Franco et al.}  

\title [mode = title]{New variable sources revealed by DECam toward the LMC: \\ the first 15 deg$^2$}  



%


\author[1,2,3]{A. Franco}[type=editor,
       orcid=0000-0002-4761-366X]
\author[1,2,3]{A. A. Nucita}
\author[1,2,3]{F. De Paolis}
\author[1,2,3]{F. Strafella}
\author[1,2,3]{S. Sacquegna}

\affiliation[1]{organization={Department of Mathematics and Physics ``E. De Giorgi'' , University of Salento, Via per Arnesano, CP-I93, I-73100, Lecce, Italy}}
\affiliation[2]{organization={INFN, Sezione di Lecce, Via per Arnesano, CP-193, I-73100, Lecce, Italy}}
\affiliation[3]{organization={INAF, Sezione di Lecce, Via per Arnesano, CP-193, I-73100, Lecce, Italy}}

\fntext[1]{Corresponding author: antonio.franco@le.infn.it}


\begin{abstract}
The Dark Energy Camera (DECam) is a sensitive, wide field instrument mounted at the prime focus of the 4 m V. Blanco Telescope in Chile. Beside its main objectives, i.e. understanding the growth and evolution of structures in the Universe, the camera offers the opportunity to observe a $\sim$ 3 deg$^2$ field of view in one single pointing and, with an adequate cadence, to identify the variable sources contained. In this paper, we present the result of a DECam observational campaign toward the LMC and give a catalogue of the observed variable sources.
We  considered all the available DECam observations of the LMC, acquired during 32 nights over a period of two years (from February 2018 to January 2020), and set up a specific pipeline for detecting and characterizing variable sources in the observed fields.
Here, we report on the first $15\deg^2$ in and around the LMC as observed by DECam, testing the capabilities of our pipeline. Since many of the observed fields cover a rather crowded region of the sky, we adopted the \isis subtraction package which, even in these conditions, can detect variables at a very low signal to noise ratio. All the potentially identified variable sources were then analyzed and each light curve tested for periodicity by using the Lomb-Scargle and \czerny algorithms. Furthermore, we classified the identified sources by using the UPSILoN neural network.
This analysis allowed us to find 70 981 variable stars, 1266 of which were previously unknown. We estimated the period of the variables and compared it with the available values in the catalogues. Moreover, for the 1266 newly detected objects, an attempted classification based on light curve analysis is presented.
\end{abstract}




\maketitle

\section{Introduction}
\label{sec-intro}

After the identification of the first variable, Mira Ceti (discovered in 1598 by Fabricius and today known as Omicron Ceti), the number of known variable stars rapidly increased, and the subsequent technological improvements, up until the implementation of cameras, made identifying variables easier.
Photography and spectroscopy allowed astronomers to classify stellar spectra with increasingly higher precision, leading to a deeper description and comprehension of the behavior of variable sources. These new studies enabled the possibility to categorize variable stars as transients, periodic, non-periodic, depending on their light curve features.

Nowadays, several classes of variable stars have been recognized and studying how their brightness varies is certainly helpful for understanding their nature and estimate their physical parameters.

Many of these variables, e.g. Cepheids and RR-Lyrae stars, are used as distance indicators thanks to their characteristic period of oscillation. In particular Cepheids can be used to measure distances up to tens of Mpc due to their higher brightness, that places them up to 6 magnitudes brighter than RR-Lyrae (\citet{LMC_var, freedman2001}). Several of these stars have been found by the MACHO collaboration (\citet{macho_micro}) and by OGLE II (\citet{ogle2}), which classified about one million of such objects in the Milky Way and in the Magellanic Clouds (\citet{ogle_onemill}).

The aim of this work is to present the results obtained by analysing repeated observations of several fields, in the direction of the Large Magellanic Cloud, obtained by the Dark Energy Camera between February 2018 and January 2020, with the main objective being the identification of variable sources. In particular, in Section \ref{sec-ana} we give some technical details on the observational survey considered in this study as well as a description of the input data and of the methods employed to reduce and calibrate the images. In section \ref{sec-identif} the main results are discussed. In particular, the detected variable stars are classified as \textit{known}, if they appear in some reference catalogues (e.g. OGLE, EROS-2, MACHO, GAIA), or \textit{new} if they do not appear in other catalogues. The new variable candidates have been further analysed using the UPSILoN software which can classify periodic variables based on their light curves. Finally, in section \ref{sec-conc} our main conclusions are discussed with a brief outlook on potential future developments.


\section{Observations and Data reduction}
\label{sec-ana}

\begin{figure}
    \centering
    {\includegraphics[width=0.45\textwidth]{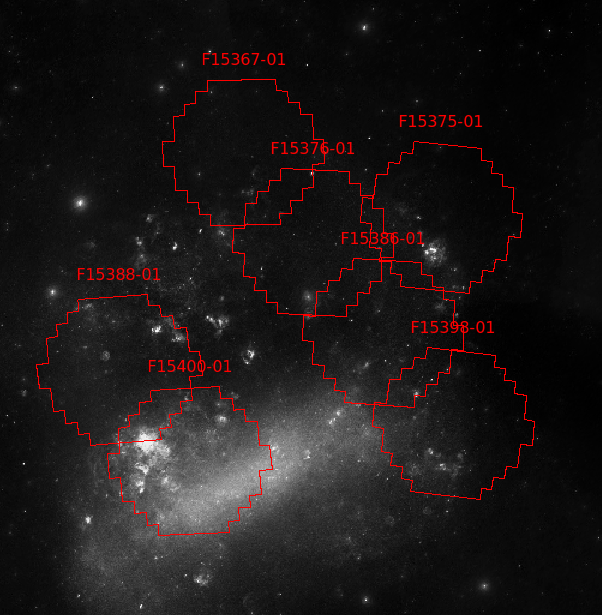}} 
    \caption{The DECam footprint for the seven fields listed in Table \ref{Table:decam_fields} studied in this work.}
    \label{Fig:footprint}
\end{figure}

The Dark Energy Camera (DECam) is  a powerful instrument mounted at the prime focus of the 4 m V. Blanco Telescope at the Cerro Tololo Inter-American Observatory (CTIO) in Chile (\citet{flaugher}). The camera consists of a grid of 62 CCDs covering a field of view (FOV) of about $3$ \degs ($\simeq 2.2$ deg wide). A two-year CTIO program (code 2018A-0273), with  Dr. William Dawson as P.I., was performed from February 2018 to January 2020, and the corresponding data were released to the public on the NOIRLab Archive website. The program was intended to intensively observe the Magellanic Clouds in order to identify potential microlensing events induced by Intermediate-Mass Black Holes, acquiring images during 32 nights over two years and observing 29 different fields of view, 23 of which pointed towards the Large Magellanic Cloud (LMC) and the other 6 towards the Small Magellanic Cloud (SMC). \footnote{This long survey was originally planned with the aim to detect microlensing events due to Intermediate-Mass Black Holes which may populate the Galactic Halo (see \citet{framp}). Here, as discussed in the previous section, we concentrate on the by-products delivered by this investigation, i.e. the detection of known and unknown variables (and their classification) in the observed fields.}
Each DECam field produced images in the SDSS (Sloan Digital Sky Surbey) bands \textit{g, r} and \textit{i} that were acquired using a band-related exposure time of $200$, $100$, and $200$ seconds, respectively. In this paper we present the results obtained analysing the first 7 DECam fields (see Table \ref{Table:decam_fields} and Figure \ref{Fig:footprint}), covering a FOV of $\sim15\deg^2$.

\begin{table*}
    \caption{For the seven analysed DECam fields towards LMC, whose technical name is indicated in the first column, the RA/DEC coordinates for each pointing are given in the second and third columns, while the acquired sky area is shown in the fourth column. The number of observations for each photometric band is given in the fifth, sixth and seventh columns. The observation dates related to the first and last acquisition are reported in the last two columns.}
    \label{Table:decam_fields}
    \centering
    \begin{tabular}{lllllllll} 
        \hline \hline
        \textbf{Field} & \textbf{RA} & \textbf{DEC} & \textbf{Sky Area} & \multicolumn{3}{c}{\textbf{N. obs}} & \textbf{DateOBS (first)} &\textbf{DateOBS (last)} \\
               & (deg) & (deg) & ($\deg^2$) & \textit{g} & \textit{r} & \textit{i}  & \textit{yyyy-mm-ddThh:mm:ss} & \textit{yyyy-mm-ddThh:mm:ss}  \\
        \hline
        F15367-01     & 80.707   & -65.134   & 2.08887    & 41   & 48   & 7 & 2018-02-18T08:34:26   & 2020-01-26T12:23:41           \\
        F15375-01     & 74.000   & -65.912   & 1.99559    & 44   & 51   & 7 & 2018-02-18T08:37:49   & 2020-01-26T12:22:56              \\
        F15376-01     & 78.336   & -66.385   & 1.42595    & 44   & 50   & 7 & 2018-02-18T08:34:34   & 2020-01-26T12:19:42             \\
        F15386-01     & 75.710   & -67.590   & 2.01091    & 44   & 51   & 7 & 2018-02-18T08:35:06   & 2020-01-26T12:19:21             \\
        F15388-01     & 85.450   & -68.050   & 2.02963    & 41   & 48   & 7 & 2018-02-18T08:39:21   & 2020-01-26T12:24:21           \\
        F15398-01     & 72.810   & -68.755   & 1.94137    & 44   & 51   & 7 & 2018-02-18T08:35:06   & 2020-01-26T12:19:04              \\
        F15400-01     & 83.000   & -69.410   & 2.06087    & 40   & 47   & 7 & 2018-02-18T08:37:27   & 2020-01-26T12:21:32             \\
        \hline \hline
    \end{tabular}
\end{table*}

In order to perform a comprehensive analysis of the images acquired during the 2018A-0273 program, a complete pipeline was written including any step necessary to produce the final stack of time sorted images.
First, we extract each extension from the mosaic images and select a reference frame to perform a first image alignment. Then, due to small errors in pointing among different epochs, we used the image astrometry to extract the common observed sky area from the different frames and crop each image within it.
After this step the brighter and saturated sources have then been masked to avoid problems during the subsequent photometric analysis. Note that, due to all these adjustments, the {\it Sky Area} value may be different for each field analysed. Moreover, possible overlaps, as shown in Figure \ref{Fig:footprint}, are not particularly relevant to increase the sampling, since observations were carried out during the same nights with a few minutes delay among the different fields. At the end, the produced images are stored as FITS files constituting the final database adopted for our analysis.

\subsection{Photometric procedure}

Because of the source crowding in the LMC fields, the photometric reduction carried out in this work was done using \textit{ISIS 2.2} (\citet{alard98, alard00}), a package that allows users to perform an accurate image difference photometric analysis. The subtraction software requires a set of parameters to produce a new sample of convolved images, highlighting luminosity changes due to intrinsic or induced source variability (see Table \ref{Table::default_config} for the default configuration parameters used for the runs). In particular, we consider a 3rd degree polynomial function for the preliminary astrometric alignment\footnote{At this step, before the astrometric alignment, we already discarded the frames suffering a bad pointing, hence having a poor overlap with the set of images (maximum 1-2 frames are discarded if necessary, but this is a rare event).
In this way, the remaining images are well matched and a subsequent alignment provided by \isis is effective to add fine tuning correction.} between frames, choosing the best signal-to-noise ratio image as common reference. After the interpolation process, the final reference image is produced by stacking the best images that appear to be clean and low noise. This is the image that, after convolution with an appropriate kernel aimed to compensate for the different seeing conditions, is subtracted to each frame of the investigated field, producing a series of subtracted images that show variable objects in the field. In this way a new image is produced enhancing the detected variable sources.
Concluding this phase, the photometric procedure provides, for each possible variable candidate, the light curve file containing the differential flux and the relative error.
The procedure finally returns a list of variable candidates with the respective pixel coordinates $X-Y$ in the image, as well as light curves with the corresponding averaged signal-to-noise ratio. The latter value is subsequently useful to discriminate actual variables from instrumental background fluctuations or saturated pixels: in our experience, if the signal-to-noise ratio estimated by \isis lies in the range $1.2-5$, then the candidates can be considered as reliable\footnote{Threshold values used for SNR were chosen after comparing detected variables with known variables for a couple of test frames, evaluating the SNR distribution for cross-matched objects. Values are then chosen from the SNR distribution as limits of correlation acceptance.}. When particularly high values are found, we check for the presence of saturated pixels, also evaluating the light curve individually.
Once the photometric analysis is completed, a period estimate is performed by using the Lomb-Scargle algorithm (\citet{lomb, scargle}), which associates a probability value to the period found for the light curve.

\begin{table}
\caption{\isis configuration parameters for the alignment, subtraction and detection processes. The first column indicates the parameter name in the script, the second column indicates the values chosen for the runs while the third one gives the relative comment.}
\begin{tabular}{lll}
\hline
\hline
\multicolumn{3}{c}{\textbf{\textit{Default\_config} file parameters}}         \\ \hline
\hline
\multicolumn{1}{l}{\textbf{Parameter}}        & \multicolumn{1}{l}{\textbf{Value}}   & \textbf{Comment}             \\ \hline
\multicolumn{1}{l}{nstamps\_x}        & \multicolumn{1}{l}{6}   & N. of stamps along  X axis                 \\ 
\multicolumn{1}{l}{nstamps\_y}        & \multicolumn{1}{l}{6}   & N. of stamps along Y axis                     \\ 
\multicolumn{1}{l}{sub\_x}            & \multicolumn{1}{l}{1}   & N. of sub-division along X axis               \\ 
\multicolumn{1}{l}{sub\_x}            & \multicolumn{1}{l}{1}   & N. of sub-division along Y axis               \\ 
\multicolumn{1}{l}{half\_mesh\_size}  & \multicolumn{1}{l}{9}   & Half kernel size                                  \\ 
\multicolumn{1}{l}{half\_stamp\_size} & \multicolumn{1}{l}{15}  & Half stamp size                                   \\ 
\multicolumn{1}{l}{deg\_bg}           & \multicolumn{1}{l}{1}   & Deg. differential bkg var         \\
\multicolumn{1}{l}{ngauss}            & \multicolumn{1}{l}{3}   & N. of Gaussians                               \\ 
\multicolumn{1}{l}{deg\_gauss1}       & \multicolumn{1}{l}{6}   & Deg. 1st Gaussian               \\ 
\multicolumn{1}{l}{deg\_gauss2}       & \multicolumn{1}{l}{4}   & Deg. 2nd Gaussian               \\ 
\multicolumn{1}{l}{deg\_gauss3}       & \multicolumn{1}{l}{3}   & Deg. 3rd Gaussian               \\ 
\multicolumn{1}{l}{sigma\_gauss1}     & \multicolumn{1}{l}{0.7} & Sigma of 1st Gaussian                             \\ 
\multicolumn{1}{l}{sigma\_gauss2}     & \multicolumn{1}{l}{2.0} & Sigma of 2nd Gaussian                             \\ 
\multicolumn{1}{l}{sigma\_gauss3}     & \multicolumn{1}{l}{4.0} & Sigma of 3rd Gaussian                             \\ 
\multicolumn{1}{l}{deg\_spatial}      & \multicolumn{1}{l}{0}   & Deg. spatial var of the Kernel           \\ \hline
\hline
\end{tabular}

\label{Table::default_config}
\end{table}

\subsection{Photometric calibration}

The difference image photometry obtained with the \isis method can be expressed in a standard magnitudes system provided we obtain a reliable photometric calibration of the reference image. To this aim DAOPHOT/ALLSTAR II (\citet{daophot}) has been used to obtain an accurate estimate of the instrumental magnitudes for point sources in the field of view of the reference image. Since the latter has been used to perform image subtraction, magnitude values for the other images could be obtained by considering the differential flux in counts provided by the subtraction process powered by \isis: in other words, the differential flux has been added to the zero-point flux, obtained via DAOPHOT analysis, of the reference image. After re-scaling all the images on the same instrumental magnitude scale, we used the Gaia EDR3 catalogue (\citet{gaiadr3}) in order to calibrate the instrumental magnitudes to the standard SDSS magnitudes, through the Gaia EDR3 photometric relationship provided by \citet{riello}. To better clarify this point, we consider the relation:
\begin{equation}
    G - m = f~(G_{BP}-G_{RP})
\end{equation}
where $G, \> G_{BP} \text{ and } G_{RP}$ are the Gaia magnitudes, $m$ is one of the Sloan magnitudes that must be calibrated and $f~$ is a polynomial function related to the Gaia color index. Depending on the SDSS magnitude, the $f~$ function parameters change in order to adapt the calibration.

Since we are faced with several single extensions (7 fields with 60 extensions each) and a complete calibration requires a substantial computational effort, we operated a few calibrations obtaining average values to be used for all fields, that is in the \textit{g} Sloan band: $g = 0.948 \> g^{\small i} + 6.053 $, where $g^{\small i}$ is the instrumental magnitude. For the same reason, the main work has been carried out analysing the \textit{g} band images and leaving the \textit{r} band images to confirm, is necessary, the nature of some interesting variables. Moreover, all frames that show either a small overlap due to inaccurate pointing or bad seeing (FWHM > 6 pixels) are discarded prior to entering them into the detection pipeline.

\begin{figure*}
    \centering	
    {\includegraphics[width=0.45\textwidth]{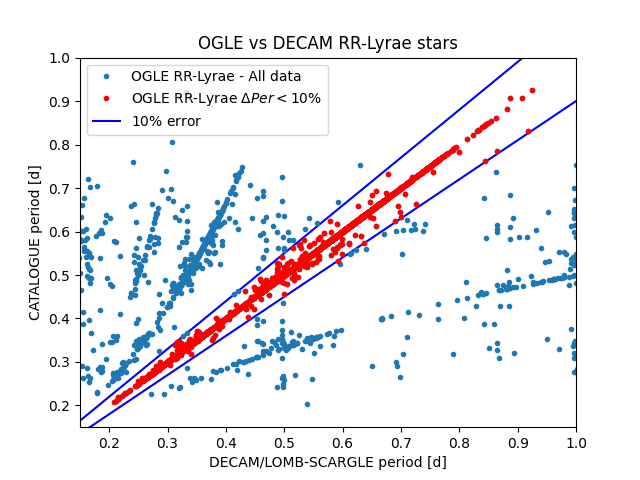}} 
    {\includegraphics[width=0.45\textwidth]{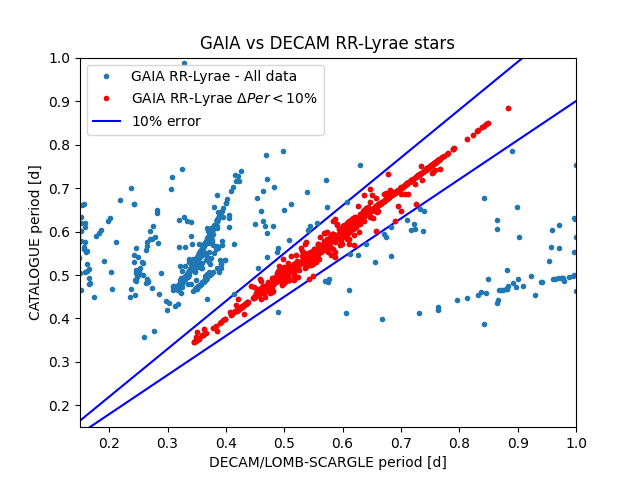}} 
    {\includegraphics[width=0.45\textwidth]{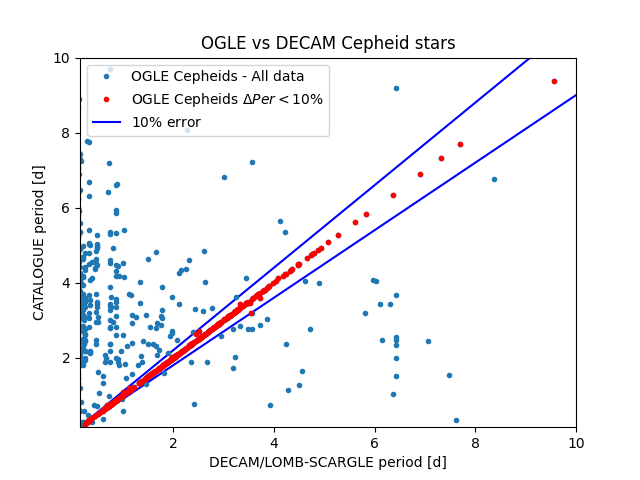}} 
    {\includegraphics[width=0.45\textwidth]{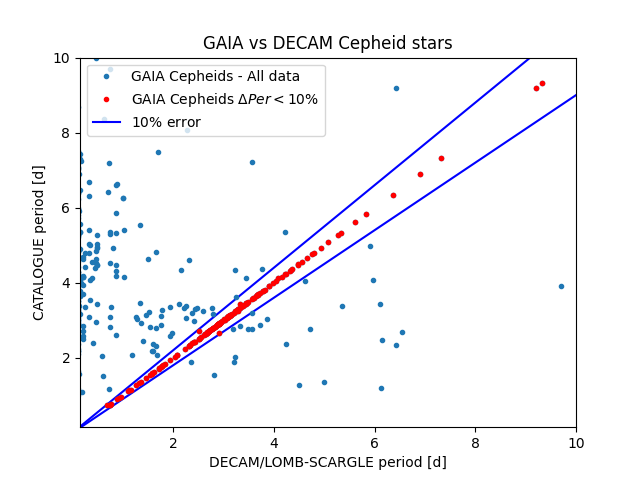}}
    \caption{The four panels show the correlation between the variable sources periods, estimated via the Lomb-Scargle periodogram, and the periods in the OGLE and GaiaDR2 catalogues (left and right panels, respectively). The two upper panels show the correlation for the RR-Lyrae stars while in the lower panels the obtained correlation for the Cepheid stars is displayed. The blue dots represent all the correlated data within the detection range of [0.1, 1.0] days for the RR-Lyrae and [0.1, 10.0] days for Cepheids, with the lower limit fixed to 0.1 days as a computational constraint. The blue lines delimit the region (red dots) in which the presently derived periods agree with previous catalogues, within 10\%.}
    \label{Fig:periods-corr}
\end{figure*}

\section{Identification of variables}
\label{sec-identif}

Variable sources are identified analysing their light curves obtained with the \isis photometric procedure. Once the variables are identified a search in the catalogues currently available in literature, which are listed in Table \ref{Table::catalogued_var}, is carried out. As a result 70 981 variables have been identified of which 69 715 are already known and 1266 appear to be previously unknown variables. In particular, in Table \ref{Table::catalogued_var}, the number of identified variables in each catalogue is reported along with the relative variable type. As one can note from the Table, most of the known variables are Cepheids, RR Lyrae, Eclipsing Binaries and Long-Period Variables. Furthermore, some variables may be found in different catalogues (e.g. the variable "OGLE LMC573.20.006559" was found in `J/AcA/59/1' [\citealt{ogle_rrl}], `J/AJ/158/16/table11' [\citealt{des_rrl}] and `I/345/rrlyrae' [\citealt{gaiadr2}]). It is clear that the total number of variables obtained by the sum of the detected sources for each catalogue is greater than 69~715 (the total number of variables detected in this work), simply because many objects may appear in more than one catalogue.

\begin{figure*}
    \centering	
    
    {\includegraphics[width=0.32\textwidth]{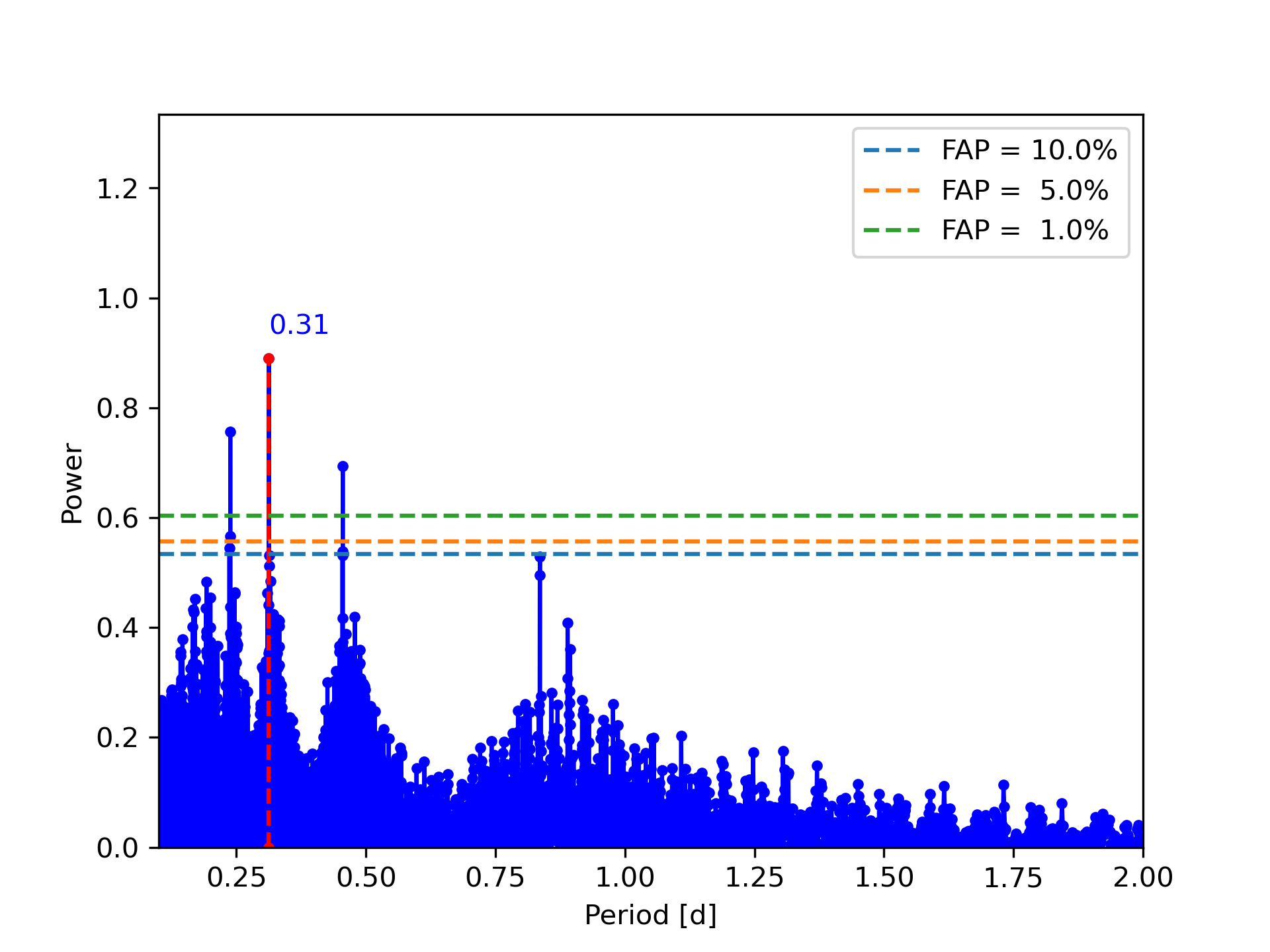}}
    {\includegraphics[width=0.32\textwidth]{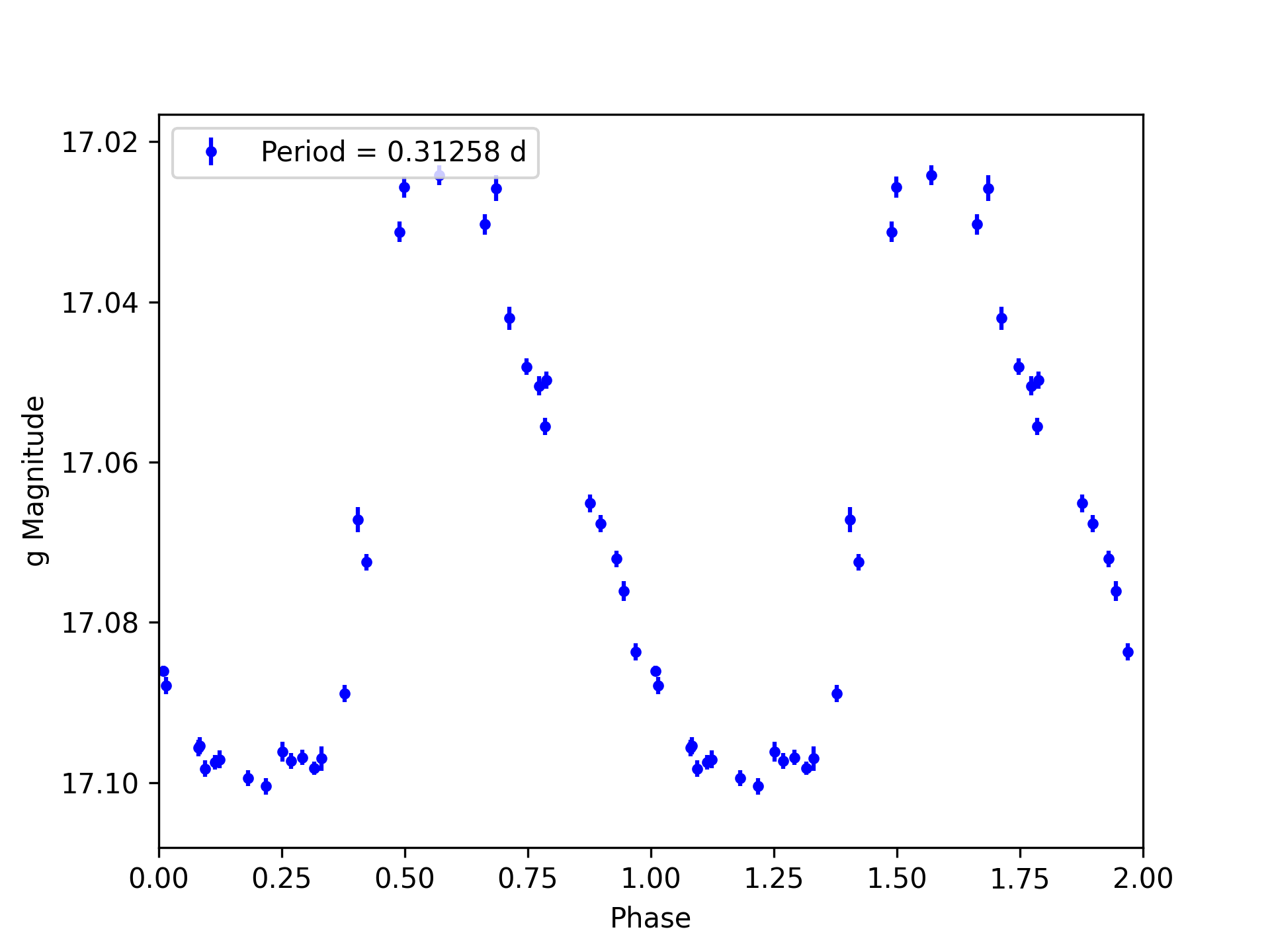}}
    {\includegraphics[width=0.32\textwidth]{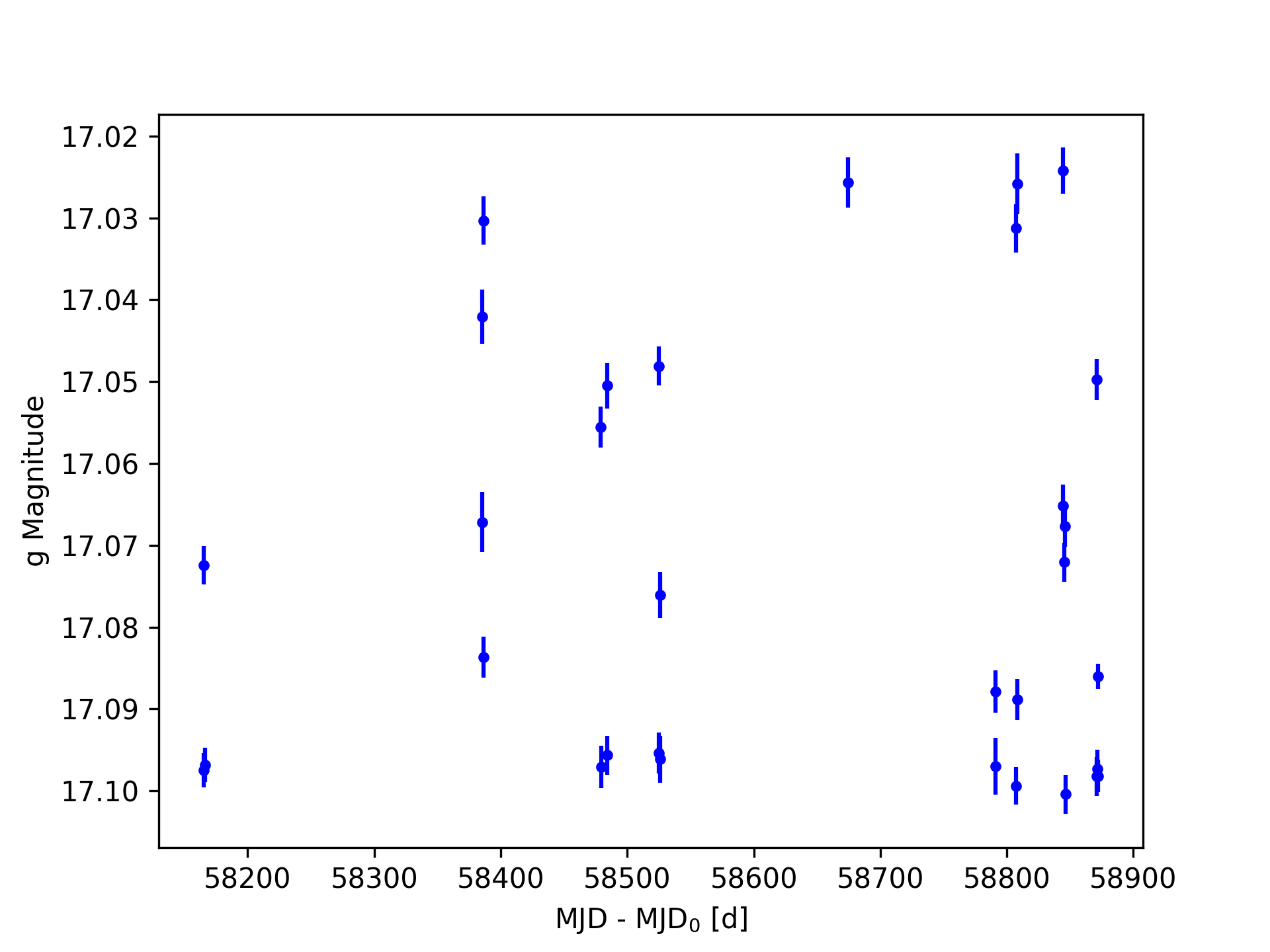}}
    {\includegraphics[width=0.32\textwidth]{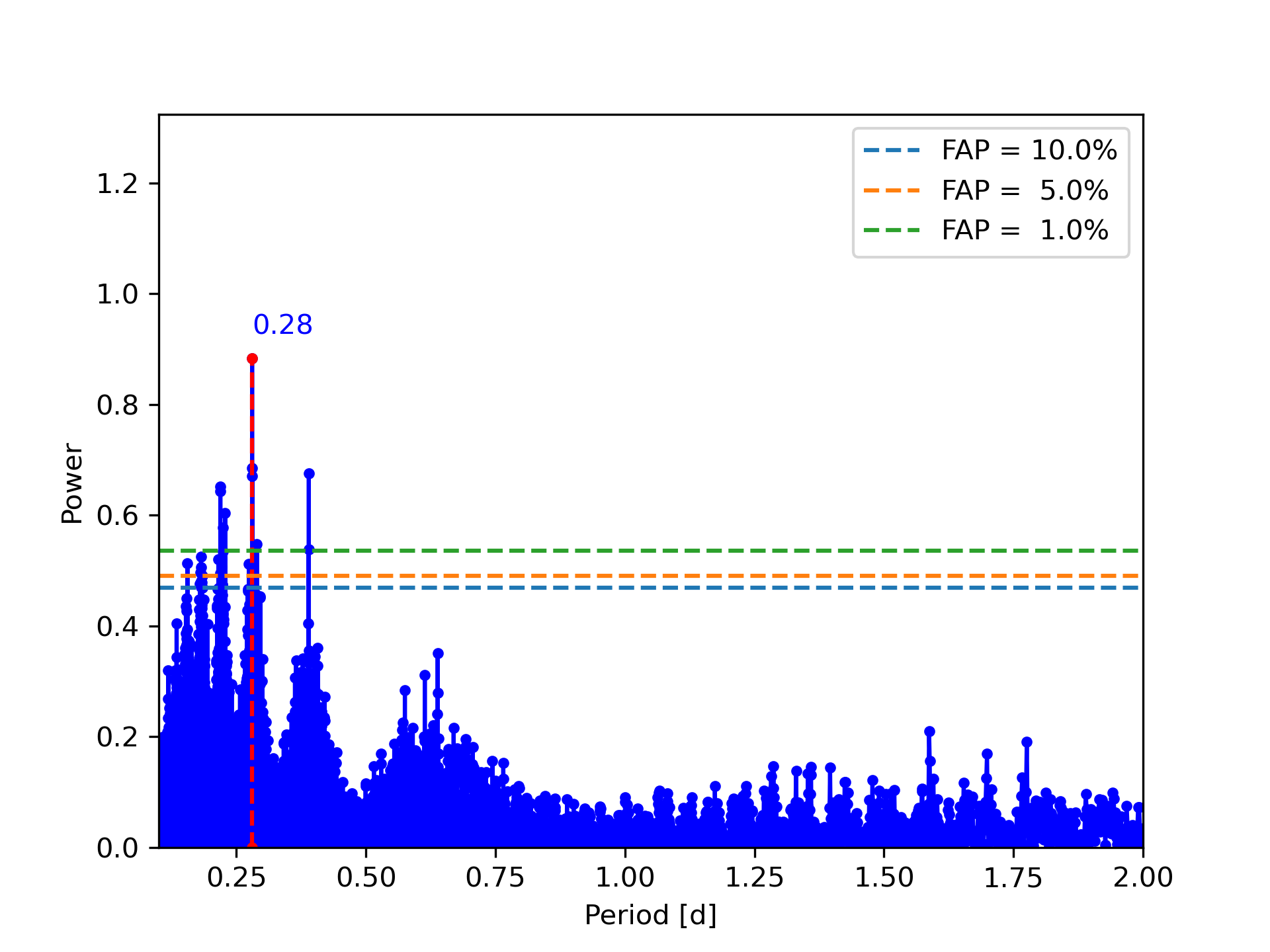}}
    {\includegraphics[width=0.32\textwidth]{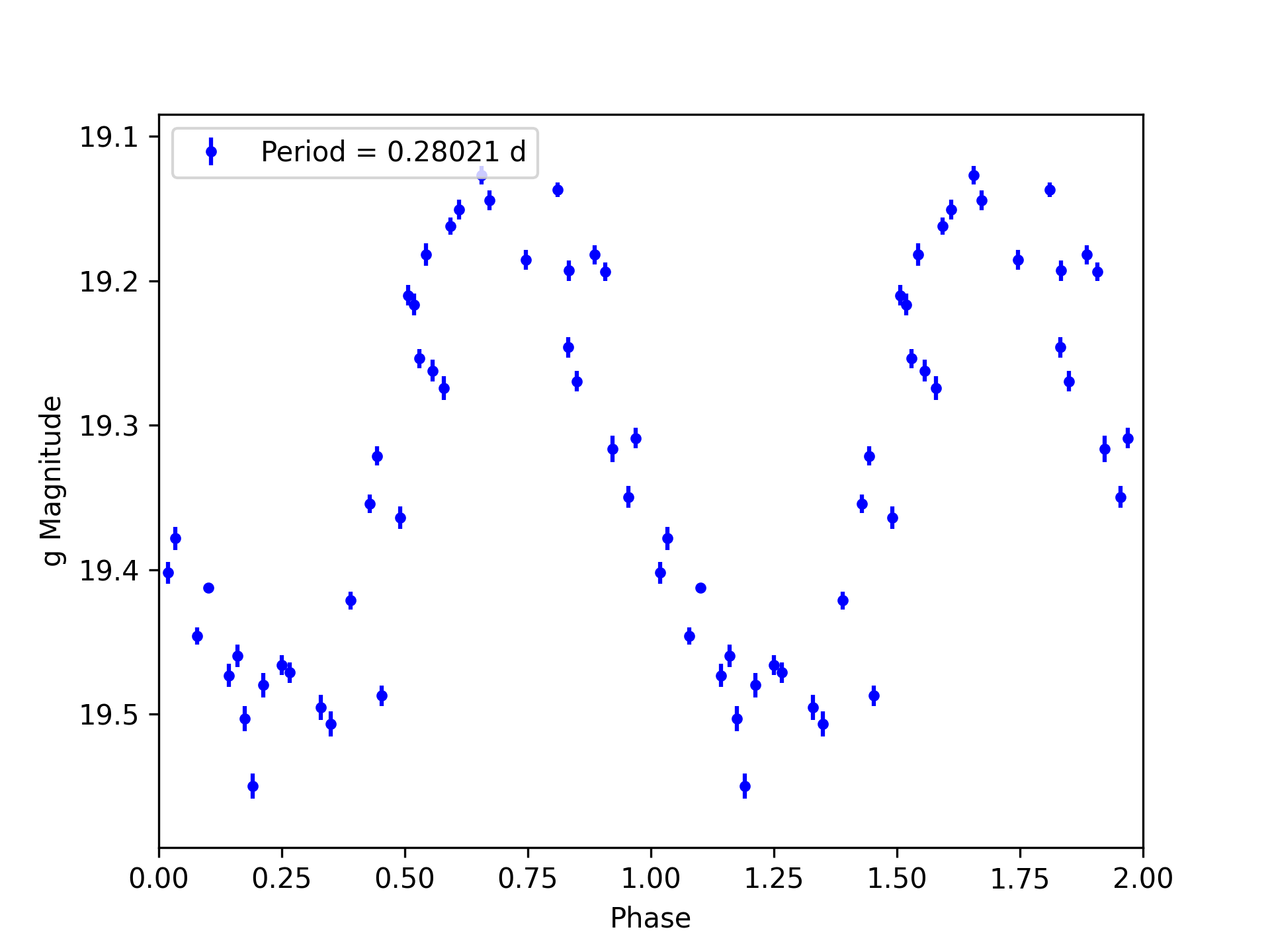}}
    {\includegraphics[width=0.32\textwidth]{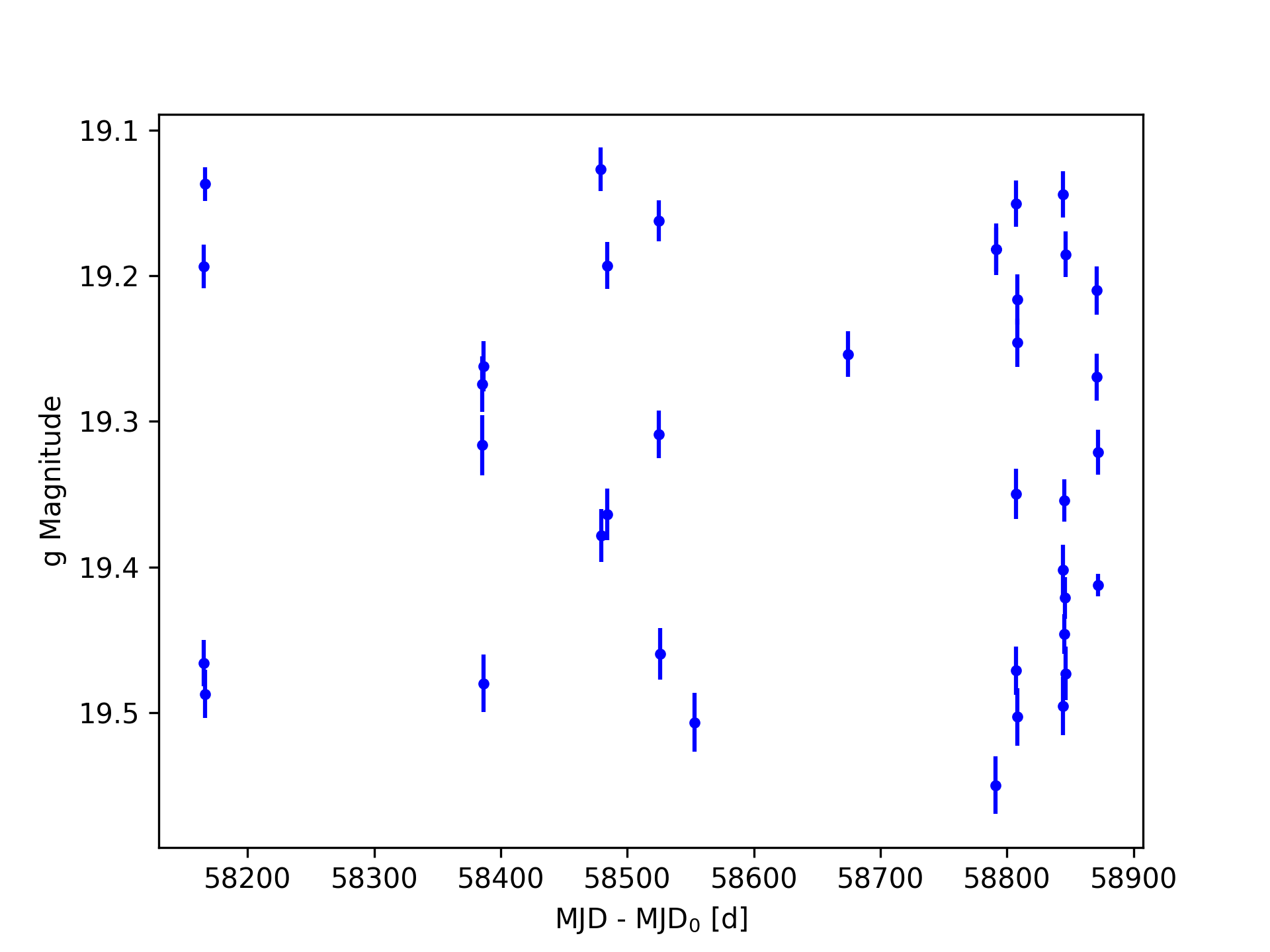}}
    {\includegraphics[width=0.32\textwidth]{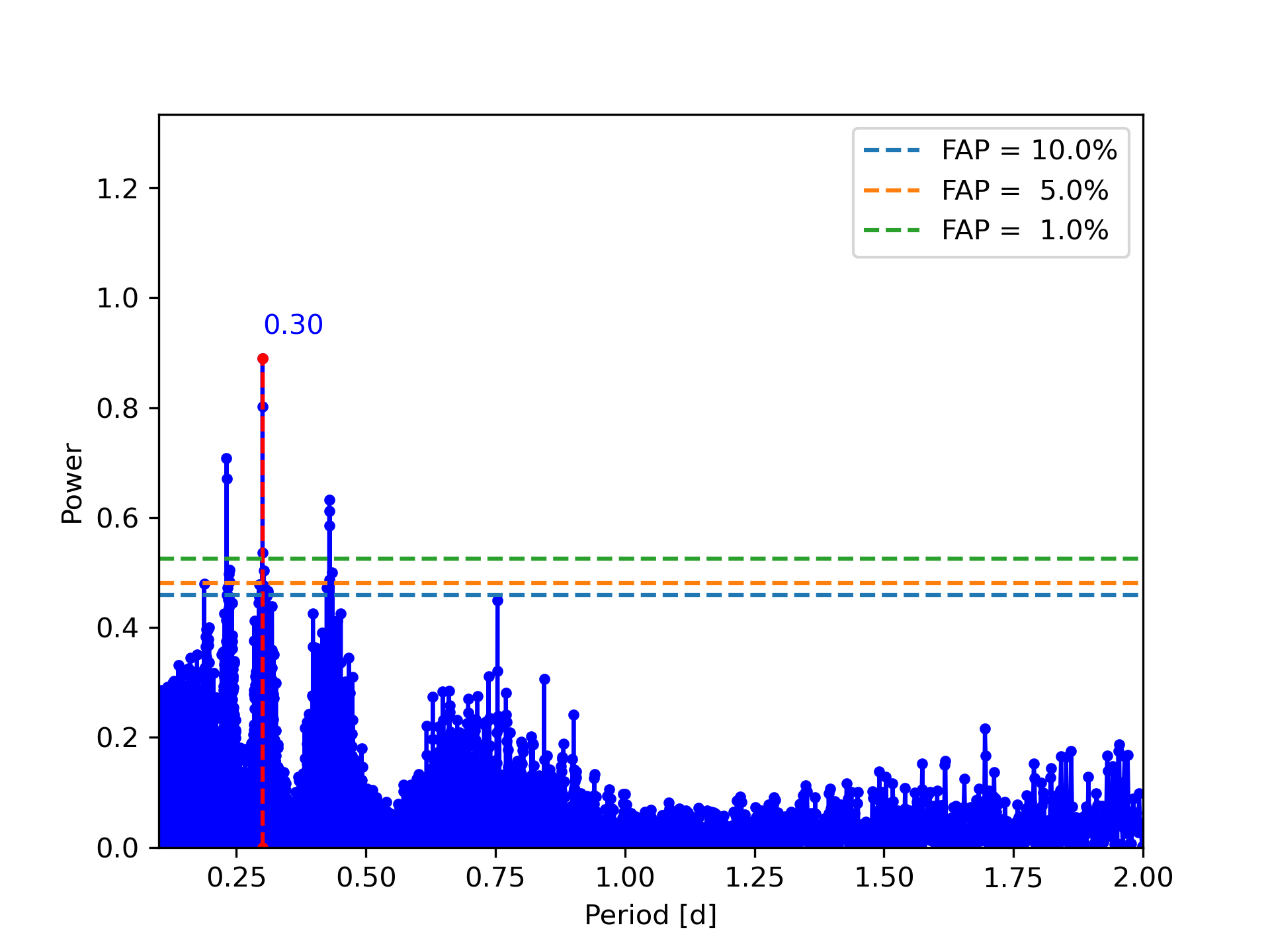}}
    {\includegraphics[width=0.32\textwidth]{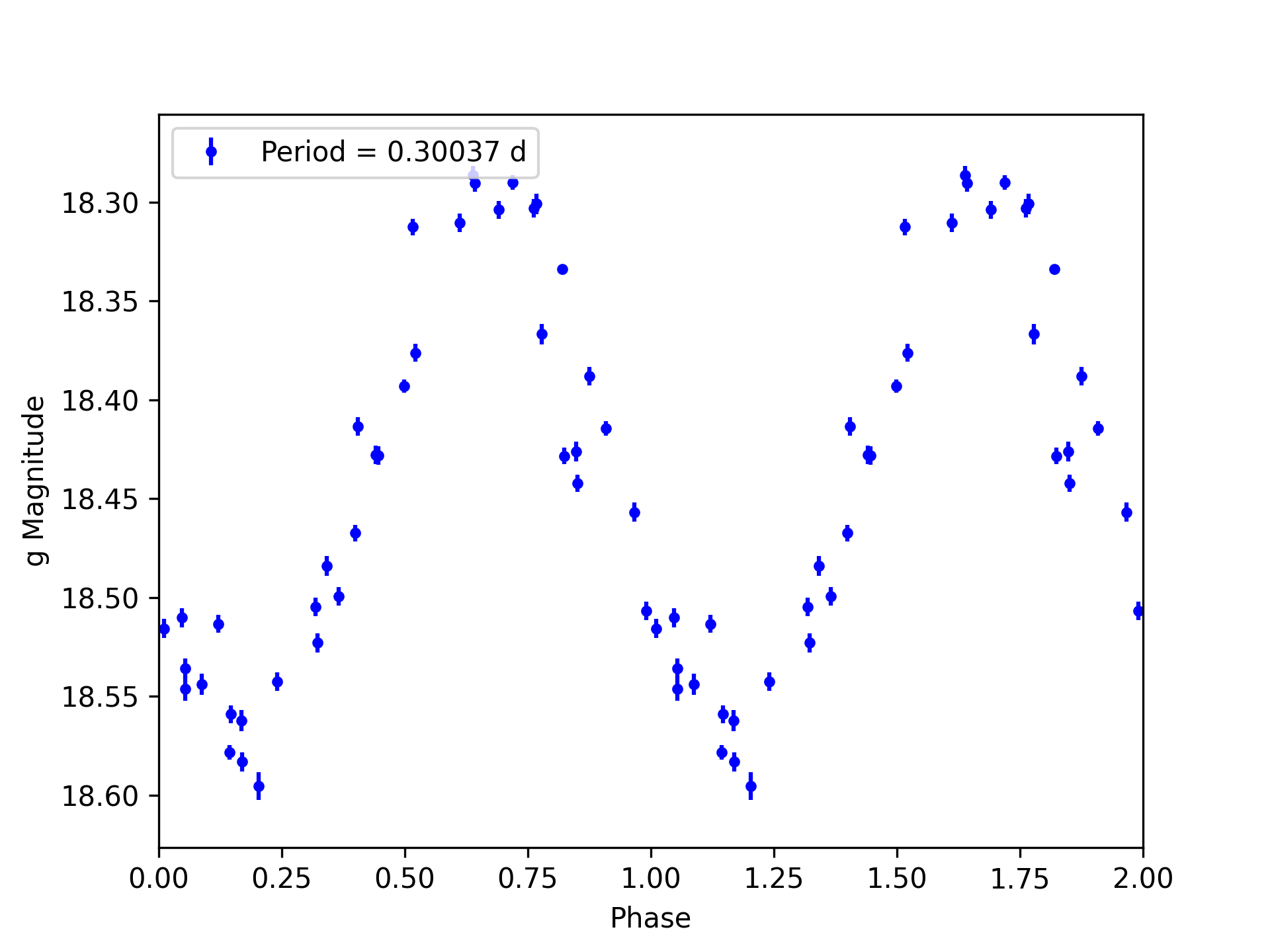}}
    {\includegraphics[width=0.32\textwidth]{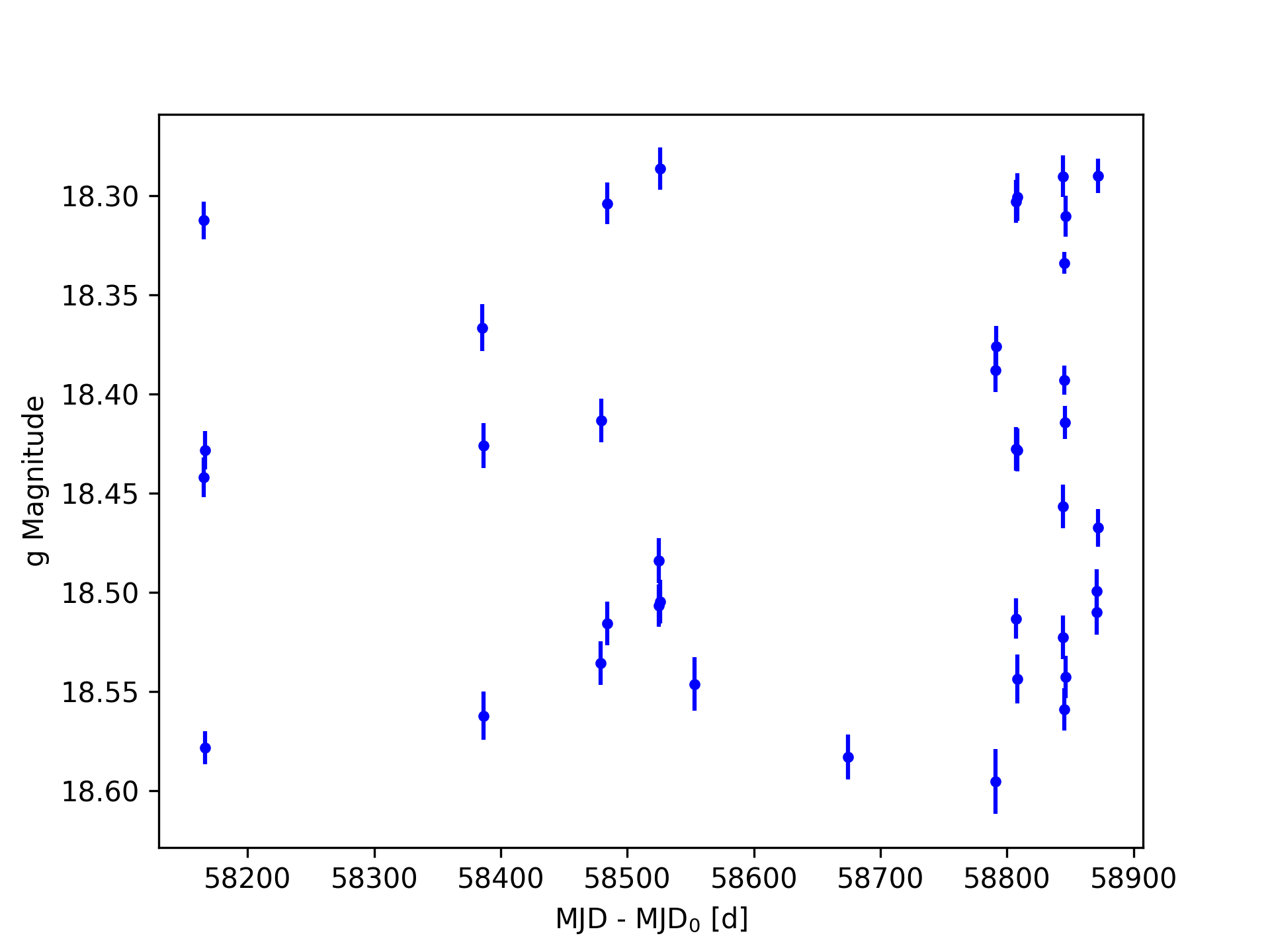}}
    
    \caption{Periodograms (left panels) , folded light curves (central panels) and unfolded light curves (right panels) for three known variable sources obtained using the Lomb-Scargle algorithm applied to the data provided by \isis. Starting from the top, the first two sources are RR-Lyrae stars (at coordinates of (82.66979, -68.83211) degrees and (72.10004, -69.47417) degrees) whose estimated period is $P^{lomb}_{RRL_1} = (0.3125794 \pm 0.000006)$ days $P^{lomb}_{RRL_2} = (0.2802124 \pm 0.000005)$ days respectively. The last one is a Classic Cepheid star (78.67071, -66.61411) with an estimated period of $P^{lomb}_{CEP} = (0.3003658 \pm 0.000005)$ days. The catalogued periods are $P^{cat}_{RRL_1} = 0.3125796$ days, $P^{cat}_{RRL_2} = 0.2802012$ days and $P^{cat}_{CEP} = 0.3003773$ days. Here the label \textit{cat} refers to the period found in the catalogues, \textit{lomb} is the period estimate returned by the algorithm and \textit{RRL} / \textit{CEP} refer to the star type, RR Lyrae or Cepheid. In the left panels, the box in the upper right refers to the false alarm probability.}
    \label{Fig:known-var-examples}
\end{figure*}

Despite the fact that the survey is more efficient for the detection of long-period variables\footnote{The survey sampling (20 -- 40 epochs in $\sim$2 years) would allow in principle a better detection of long-period variables, since short periods might require a denser sampling for their detection.}, the Lomb-Scargle periodogram is able, due to the good accuracy of the \isis photometry, to identify many short period variables with a period shorter than 20 days (about 90\% of the total variables).
In relation to this, in Figure \ref{Fig:periods-corr} we present the correlations (blue dots) between the periods found via Lomb-Scargle analysis and the values from the catalogues of OGLE (left panels) and GAIA DR2 (right panels), for both RR-Lyrae (top panels) and Cepheids stars (bottom panels). The blue lines in the figure delimit the region (red dots) in which the periods obtained in this work agree, within 10\%, with those found in previously published catalogues.
For about 70\% among the over 4800 considered RR-Lyrae and around 45\% among about 800 Cepheids stars the correlation is clear. The estimated periods are in the range of the Lomb-Scargle period limits, i.e. $0.1 - 200$ days.
Three examples of periodic analysis are shown in Figure \ref{Fig:known-var-examples}, reporting the periodogram, the phased light curve  and the raw data, spanning the whole 2 years observation window, for two RR-Lyrae stars (from the top, the first two couples of panels) and one Cepheid variable (the last one). In particular, in the periodogram panels, dashed horizontal lines indicate the threshold levels corresponding to the False Alarm Probability (FAP) of 0.1 (blue), 0.05 (orange) and 0.01 (green). A vertical red line indicates the peak of the power spectrum and, therefore, the most probable period. This period is then used to obtain the corresponding phased light curve, displayed in the right panels in Figure \ref{Fig:known-var-examples}.

We would also like to stress the various patterns that arise for the RR-Lyrae comparison in Figure \ref{Fig:periods-corr}. In fact, noisy observations lead to incorrect estimates of the peak in the periodogram, corresponding to false frequencies, therefore returning a wrong estimate of the period. This effect may be due either to the interaction between the window function and the underlying spectral power or to the selection of a frequency related to a harmonic that, in the periodogram, shows a higher power with respect to the true frequency. Both kinds of failure may induce spurious effects, as shown in Figure \ref{Fig:periods-corr}. Hence, assuming that the available catalogue reports the true binary periodicity, we select the diagonal path only, discarding the other patterns. An exhaustive discussion of the behaviour of the Lomb-Scargle algorithm in such cases can be found in \citet{lombscargle2}.

It is worth noting that, in spite of the high number of known variables identified, the data obtained by the present survey do not allow the identification of the whole number of variables already known due to some technical limits. For example, very bright and saturated objects are masked and discarded a priori so that possible variables associated are therefore not found. Furthermore, since the survey sampling is not homogeneus, some variables can be missed due to their low SNR (lower than the confidence range of 1.2 -- 5 considered above) that arises in such cases. As a consequence, these variables may be erroneously identified as background fluctuations and discarded after the analysis.

\begin{table*}
    \caption{List of the catalogues used to perform the correlation with the known variables. The first column indicates the catalogue name; in the second column the variable type is given and in the third column the number of the identified variables is reported. The last column indicates the reference in literature.}
    \centering
    \begin{tabular}{llll}
    \hline
    \hline
    \textbf{Catalogue}  & \textbf{Type}                  & \textbf{Number}  & \textbf{Reference}  \\ 
    \hline
    J/AcA/60/1          & $\delta$ Scuti stars           & 567                   & \citet{ogle_delta}\\ 
    J/AcA/60/179        & Double Periodic Variables      & 31                   & \citet{ogle_double} \\ 
    J/AcA/58/293        & TypeII and Anom. Cepheids      & 47                  & \citet{ogle_cep2} \\ 
    J/AcA/58/163        & Classical Cepheids             & 793                 & \citet{ogle_cep} \\ 
    J/AcA/61/103        & Eclipsing Binaries             & 6302                & \citet{ogle_eb} \\ 
    J/AcA/59/1          & RR Lyrae stars                 & 4849                & \citet{ogle_rrl} \\ 
    J/AcA/59/239        & Long-Period Variables          & 20760               & \citet{ogle_lpv} \\ 
    J/AcA/59/335        & R Coronae Borealis stars       & 4                   & \citet{ogle_rcb} \\ 
    J/AJ/158/16/table11 & RR Lyrae stars                 & 226                 & \citet{des_rrl} \\ 
    J/AcA/66/421/ecl    & Eclipsing Binaries             & 8732                & \citet{ogle_eb2} \\
    I/345/cepheid       & Cepheids stars                 & 442                 & \citet{gaiadr2} \\ 
    I/345/lpv           & Long-Period Variables          & 1048                & \citet{gaiadr2} \\ 
    I/345/rrlyrae       & RR Lyrae stars                 & 2701                & \citet{gaiadr2} \\ 
    J/ApJ/663/249       & Eclipsing Binaries             & 231                 & \citet{macho_eb} \\ 
    J/AJ/136/1242       & Long-Period Variables          & 9250                & \citet{macho_lpv} \\
    J/A+A/566/A43       & Periodic Variable stars        & 20300               & \citet{macho_var} \\ 
    II/247              & Variable Stars                 & 1050                & \citet{eros_var} \\ 
    J/A+A/536/A60       & Long-Period Variables          & 4652                & \citet{eros2_lpv} \\ 
    I/358/vcep          & Cepheid stars                  & 544                 & \citet{gaiadr3_2022} \\ 
    I/358/vlpv          & Long-Period Variables          & 3272                & \citet{gaiadr3_2022} \\ 
    I/358/vrrlyr        & RR Lyrae stars                 & 3436                & \citet{gaiadr3_2022} \\ 
    J/MNRAS/424/1807    & Cepheid stars                  & 185
    & \citet{vista_cep} \\ 
    J/MNRAS/443/432     & Eclipsing Binaries             & 200
    & \citet{vista_ecl} \\ 
    \hline
    \hline
    \end{tabular}
    \label{Table::catalogued_var}   
\end{table*}

\subsection{New variable stars}

\begin{figure}
    \centering	
    \includegraphics[width=0.45\textwidth]{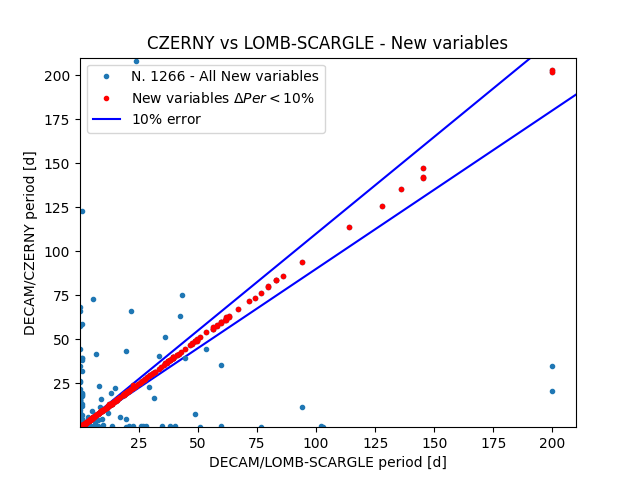}
    \caption{Period-Period diagram showing the correlation between the new variable sources periods found with the Lomb-Scargle (horizontal axis) and \czerny (vertical axis) algorithms (blue dots). The \textit{x} and \textit{y} axes are constrained in the range $0.1 - 200$ days in order to avoid wrong period estimates of sources with period close to the observational window. The lower limit of 0.1 days comes from the computational limit appropriate for both algorithms. We select only objects whose detected period found with the Lomb-Scargle periodogram differs within 10\% from the period found with the \czerny algorithm. In this way we found 634 over 1266 variables that satisfy our criteria, that is about 50\% of the previously unknown variables.}
    \label{Fig:newvar-periods-corr}
\end{figure}

In addition to the 69~715 known variables, 1266 are classified as new variable stars, either periodic or non-periodic.
The comparison of the periods provided by the Lomb-Scargle and \czerny (\citet{czerny1, czerny2}) algorithms is shown in Figure \ref{Fig:newvar-periods-corr}, where comparable results can be seen in $\sim$ 50\% of cases.

As already done for the known variables, we selected only stars whose percentage period difference, computed in this case between the period obtained with the Lomb-Scargle periodogram and the \czerny algorithm, is less than 10\%. Moreover, the search of the period has been similarly performed in the range of $0.1 - 200$ days, the latter value representing the third part of the observational window adopted to avoid border effects near the edges of the window. Since we can not always correlate the two periods obtained with the two algorithms within the same periods range, many variables seem to have different period estimates: these may sometimes be large enough for them not to appear in Figure \ref{Fig:newvar-periods-corr}. The Figure shows 634 correlated sources (red dots) over the whole sample of 1266 previously unknown variables (blue dots), corresponding to the 50\% of cases. The blue lines indicate the 10\% limits used to evaluate the correlation.

\begin{figure*}
    \centering	
    \includegraphics[width=0.9\textwidth]{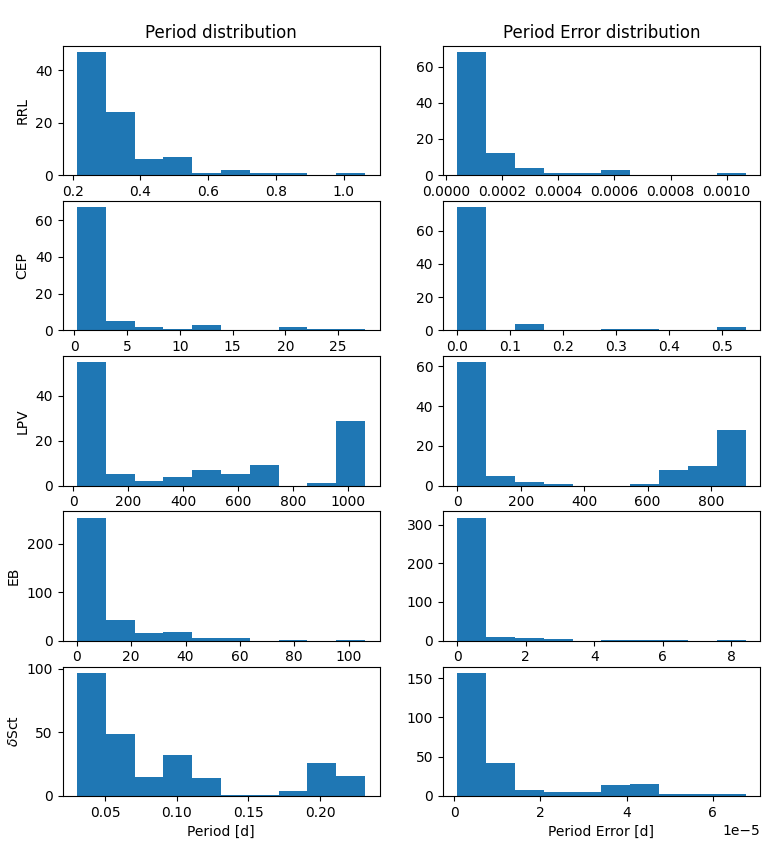}
    \caption{Periods (left panels) and associated errors (right panels) histograms for the new variables. From the top, each couple of histograms refer to a superclass (see Table \ref{Table::var_classes}) of variables, labeled on the left as returned by USPILoN, with the exception of the class {\it CEP} that refers to both {\it CEPH} and {\it T2CEPH}.}
    \label{Fig:NewVAR_PerErrHist}
\end{figure*}

\begin{figure*}
    \centering	
    {\includegraphics[width=0.45\textwidth]{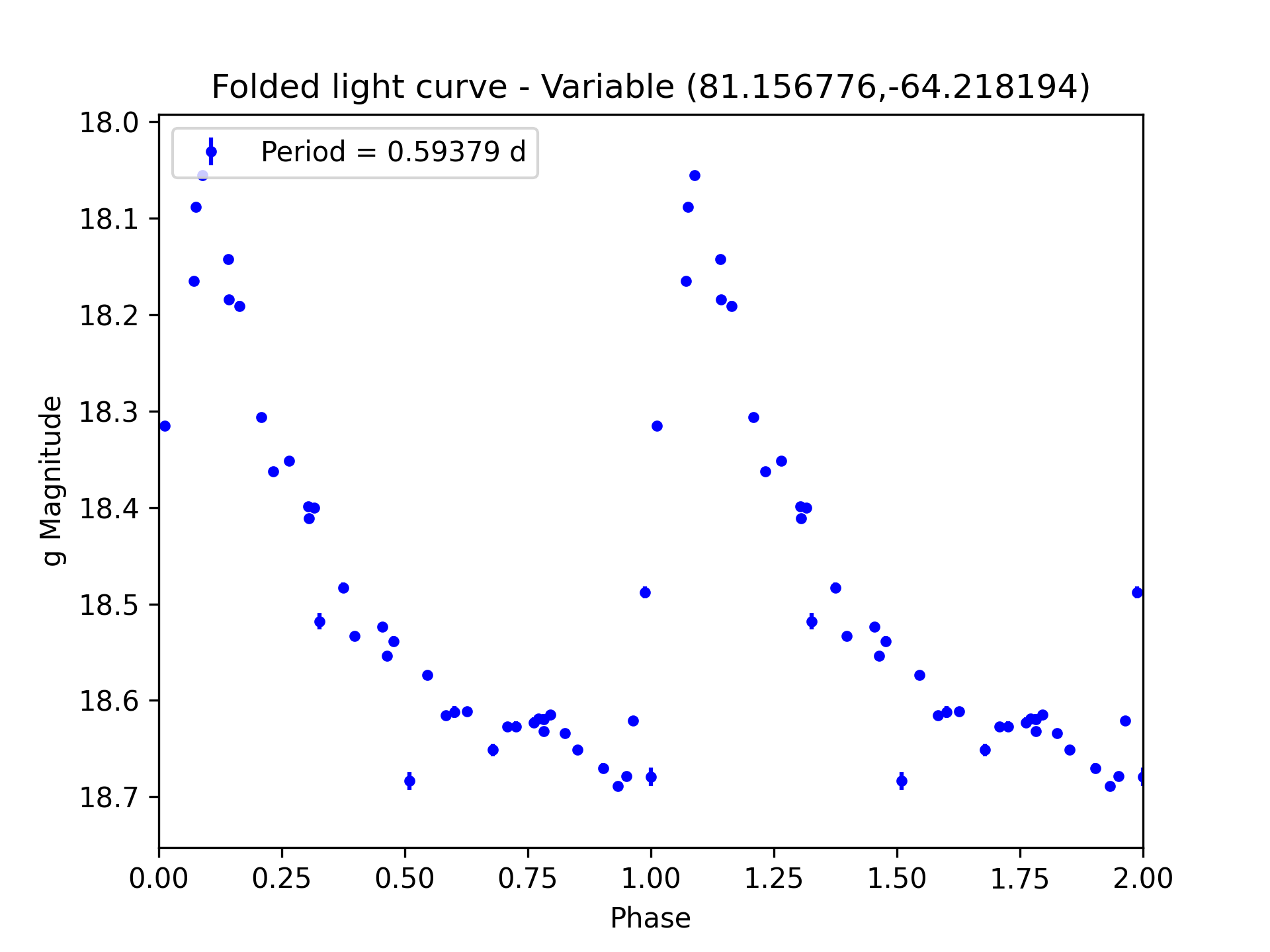}}
    {\includegraphics[width=0.45\textwidth]{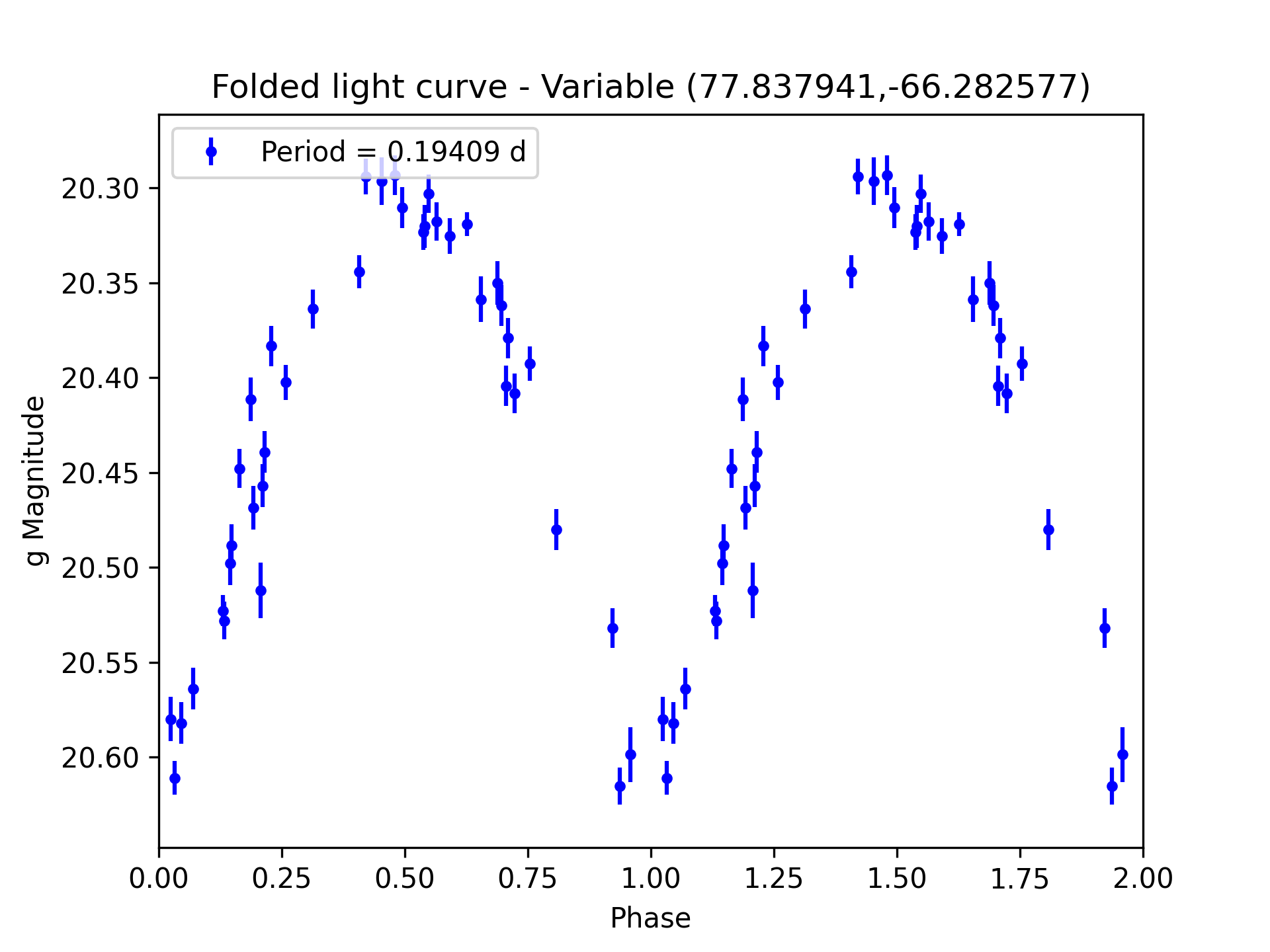}}
    {\includegraphics[width=0.45\textwidth]{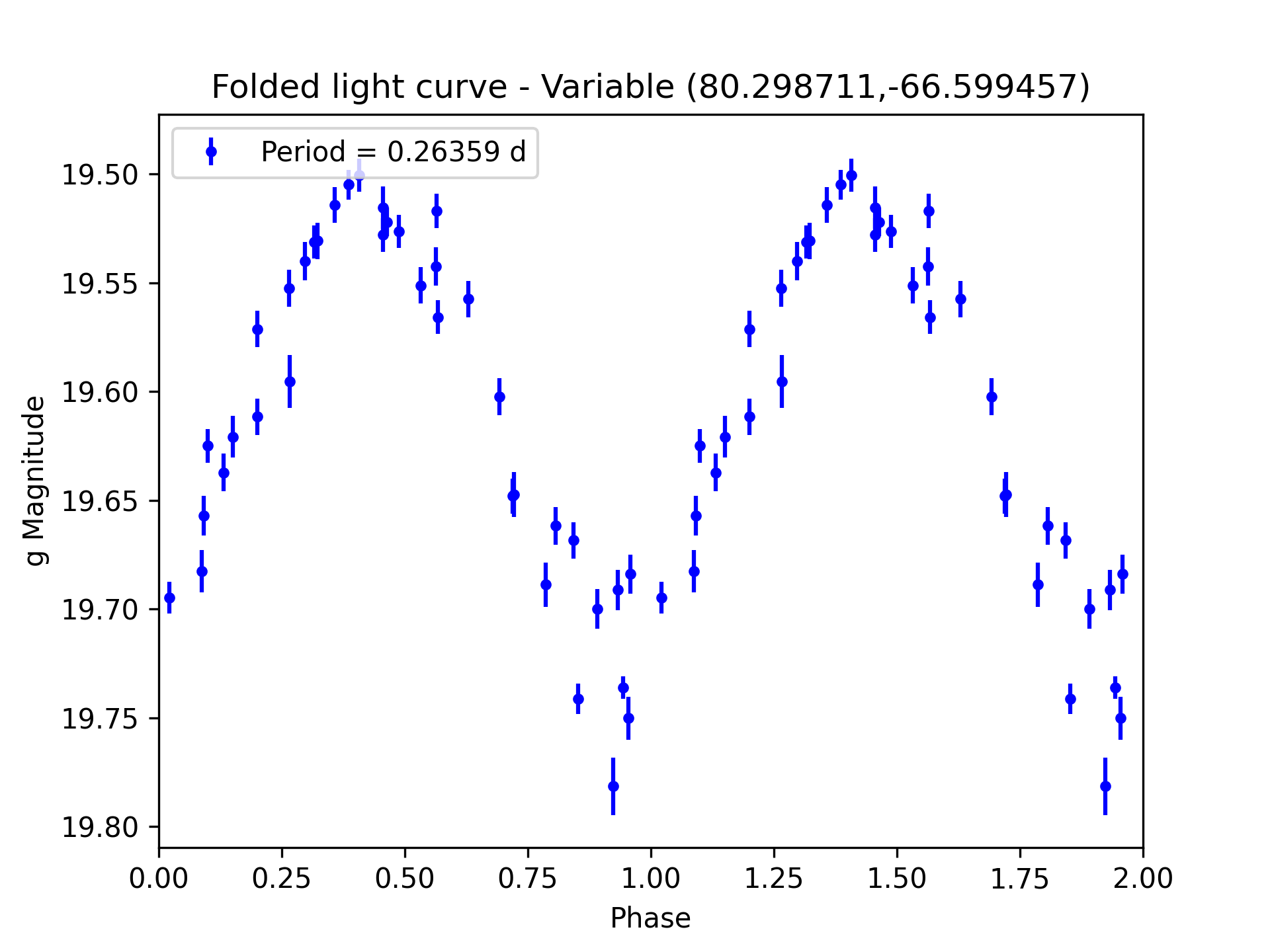}}
    {\includegraphics[width=0.45\textwidth]{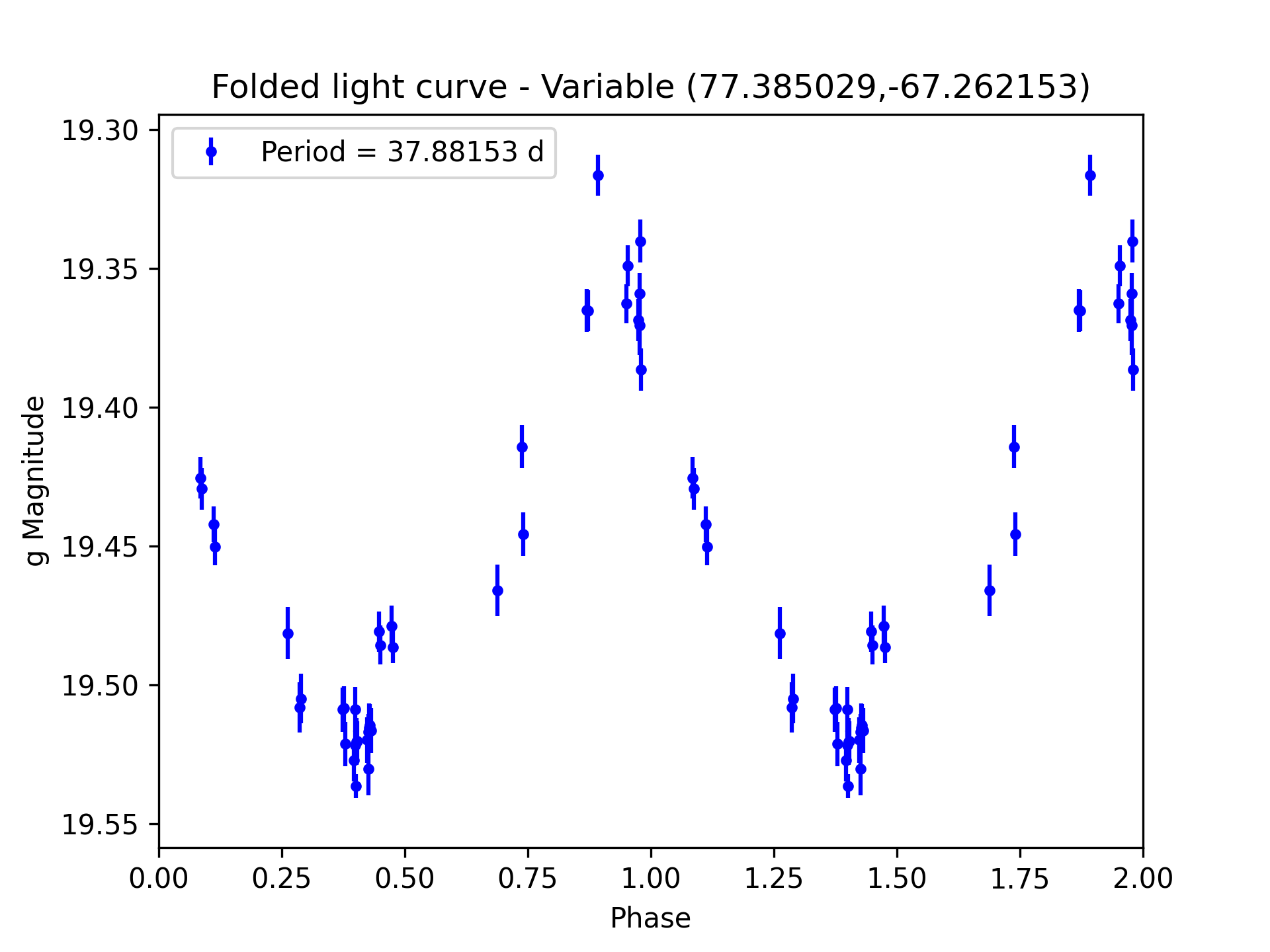}}
    \caption{Phased light curves for four new variable sources obtained using the period estimated by the UPSILoN neural network. Starting from top left, clockwise, the estimated periods are $P_1 = (0.5938 \pm 0.0003)$ days, $P_2 = (0.19 \pm 3\times10^{-5})$ days, $P_3 = (0.26 \pm 6\times10^{-5})$ days and $P_4 = (37.88 \pm 1.75)$ days. The UPSILoN software classifies these four objects as RRL\textunderscore ab, DSCT, CEP\textunderscore 1O and LPV\textunderscore SRV\textunderscore AGB\textunderscore O.}
    \label{Fig:new-var-examples}
\end{figure*}

The classification of these variables has been carried out by using the UPSILoN neural network (\citet{upsilon}). This software is able to classify periodic variable sources observed by any survey with an arbitrary time sampling using OGLE and EROS-2 surveys products as training data. As a result, a possible classification for each candidate variable is returned with an associated probability parameter. A preliminary test on the classification quality provided by UPSILoN has been carried out for a sample of known variables. Results show a good correlation in about 57\% of cases between the estimated class and the actual one. Missed or mismatched correlations are connected to the sampling and the phased light curve produced after the analysis: due to the relatively low number of epochs (20--40 epochs in $\sim$2 years), folding may produce light curves sometimes similar to each other, even though they correspond to objects of different classes (e.g. more or less sinusoidal than expected). This scenario may lead to an incorrect classification despite the correct estimate of the associated period. Table \ref{Table::var_classes} shows the number of new detected variable objects associated to each respective class. As an example, in Figure \ref{Fig:new-var-examples} are displayed four light curves that refer to new variables detected. The four examples are indicative of the possibility to detect both long and short period variables, as shown by the returned period estimates which are of about 0.59, 0.19, 0.21 and 37.88 days, respectively. Moreover, a possible classification has been provided, recognizing the probable RR Lyrae, $\delta$Scuti, Cepheid, Eclipsing Binary and Long-Period Variable nature for these objects. In Figure \ref{Fig:NewVAR_PerErrHist} periods (left panels) and associated errors (right panels) histograms for each class of variables are reported.

A fraction of these variables has been flagged as \textit{NonVar} by UPSILoN. However, despite this classification, we avoided to discard them a priori since the \isis software suggests an intrinsic variability, even if at a low amplitude level, that can be seen by-eye in the light curves. Such cases might be defined as "wrong" {\it NonVar}, due to the inability of UPSILoN in recognising low amplitude variables.
In addition, variables labeled as \textit{Unknown} in the Table \ref{Table::var_classes} are objects for which the algorithm does not reach any convergence, and therefore no information can be obtained.

\subsection{Cepheids Period-Luminosity relation}
\begin{figure*}
    \centering	
    {\includegraphics[width=0.45\textwidth]{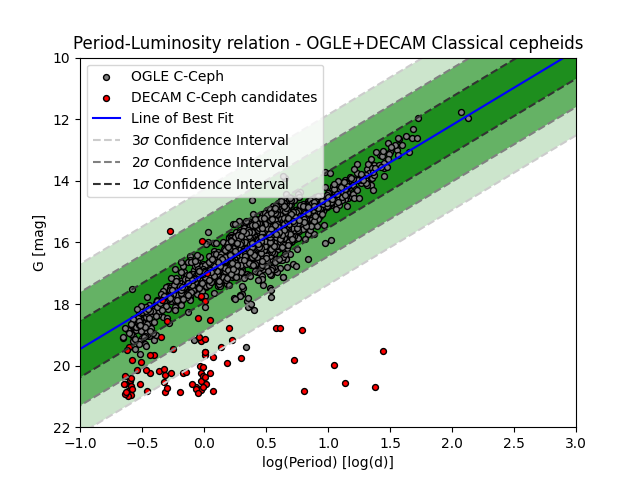}}
    {\includegraphics[width=0.45\textwidth]{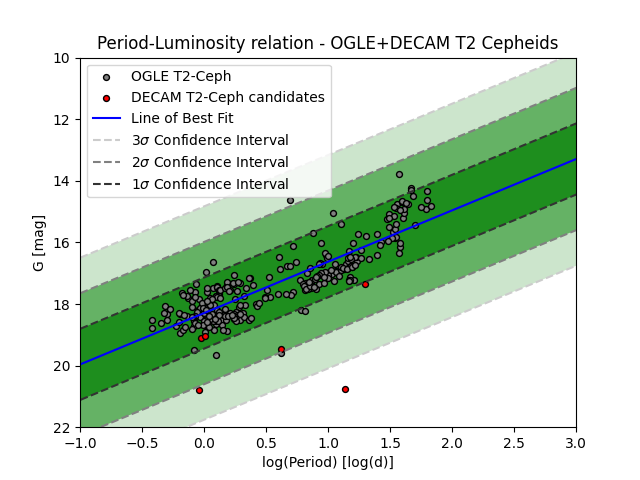}}  
    \caption{Period-Luminosity relation for Classical Cepheids (left panel) and Type-II Cepheids (right panel). The figures report the logarithm of the period extracted for the OGLE data (grey dots) and the DECAM candidates (red dots) over the \textit{x} axis and the GAIA G magnitude over the \textit{y} axis. The different confidence levels are showed as different shades of green, where the darkest one refers to the 1$\sigma$ level.}
    \label{Fig:lp-relations}
\end{figure*}

Since a fraction of the new variables has been classified as Cepheid, we use the Period-Luminosity relation in order to discriminate actual Cepheid candidates from spurious results, based on the results provided by the neural network. We consider the OGLE Cepheids period (\citealt{ogle_cep}) and the corresponding GAIA DR3 (\citealt{gaiadr3_2022}) magnitudes, if available, as reference data, obtaining a linear regression model that provides the constrains to our candidates. Figures \ref{Fig:lp-relations} show the Period-Luminosity relation for Classical (left panel) and Type-II (right panel) Cepheids, with OGLE (grey) and DECAM (red) data. Classical Cepheids include the {\it CEPH\_1O}, {\it CEPH\_F} and {\it CEPH\_Other} UPSILoN classes while the remaining {\it T2CEPH} refers to the Type-II Cepheids. Results for Classical and T2 Cepheids show that about 54\% (49 Classical and 4 T2 Cepheids) of the all DECAM Cepheids candidates agree within $3\sigma$ with the OGLE reference data. The last column of the Table \ref{Table::samplecatnewvar} accounts for the goodness of these candidates, indicating with a binary flag (0 or 1) when the period and the magnitude of the Cepheid candidate are in agreement with the Period-Luminosity relation of the reference data. In the latter cases the flag equals 1, while in all other cases is 0.

\begin{table}
\caption{Tentative classification based on the UPSILoN software for the 1266 new variables detected within the 15 \degs towards LMC. \textit{NonVar} and \textit{Unknown} targets refer to unclassified objects, either non variables or variables with an unclear variability class. An exhaustive explanation of the nomenclature adopted by the UPSILoN's authors can be found in \citet{upsilon}.}
\begin{tabular}{lll}
\hline
\hline
\textbf{Superclass} & \textbf{Subclass} & \textbf{Number} \\ \hline
DSCT        &               &   314         \\
RRL         &               &               \\
            &   ab          &   45          \\
            &   c           &   33          \\
            &   d           &   2           \\
            &   e           &   36          \\
CEPH        &               &               \\
            &   F           &   16          \\
            &  1O           &   53          \\
            & Other         &   20          \\
EB          &               &               \\
            &   EC          &   110         \\
            &   ED          &   59          \\
            &   ESD         &   244         \\
LPV         &               &               \\
            &   Mira AGB C  &   6           \\
            &   Mira AGB O  &   3           \\
            &   OSARG AGB   &   24          \\
            &   OSARG RGB   &   17          \\
            &   SRV AGB C   &   43          \\
            &   SRV AGB O   &   47          \\
T2CEPH      &               &   9           \\
NonVar      &               &   142         \\
UNKNOWN     &               &   43          \\ \hline
{\bf Total} &               &   {\bf 1266}  \\
\hline
\hline             
\end{tabular}
\label{Table::var_classes}
\end{table}

\subsection{The DECam catalogue}

We present the results of this work in two different catalogues, which contain the information collected on both the known and the new variables. In particular, the first catalogue holds 69 715 entries, listing the known variables detected. The catalogue contains positional information (Right Ascension and Declination in the ICRS), the period found with the Lomb-Scargle algorithm together with the associated error and the power in the periodogram spectrum, the period found with the \czerny algorithm and the periods found in some reference catalogues (listed in Table \ref{Table::catalogued_var}). All the column names are reported in Table \ref{Table::knowncatnames} along with the associated comments.

\begin{table}
\caption{Data columns stored in the known variables catalogue. RA\_ICRS and DE\_ICRS are the coordinates, P1 is the \czerny estimated period, P2 is the Lomb-Scargle estimated period with eP2 its error and powP2 the power associated to the peak in the periodogram. From PerCat1 to PerCat23, the known period found in literature is reported for each catalogue used in this work for the correlation phase.}
\begin{tabular}{lll}
\hline
\hline
\textbf{Column Name}  & \textbf{Comments}  & \textbf{Unit}                  \\ 
\hline
RA\_ICRS   &   Right Ascension  &  deg \\
DE\_ICRS   &   Declination   &   deg \\
P1        &     \czerny Period  &   days \\
P2        &     Lomb-Scargle Period &   days \\
eP2        &    Error on Lomb-Scargle Period    &   days \\
powP2        & Periodogram power &    \\
PerCat1      & \citet{ogle_delta} Period &   days \\
PerCat2      & \citet{ogle_double} Period &   days \\
PerCat3      & \citet{ogle_cep2} Period &    days\\
PerCat4      & \citet{ogle_cep} Period &    days\\
PerCat5      & \citet{ogle_eb} Period &    days\\
PerCat6      & \citet{ogle_rrl} Period &    days\\
PerCat7      & \citet{ogle_lpv} Period &     days\\
PerCat8      & \citet{ogle_rcb} Period &    days\\
PerCat9      & \citet{des_rrl} Period &    days\\
PerCat10     & \citet{ogle_eb2} Period &    days\\
PerCat11    & \citet{gaiadr2} Period &    days\\
PerCat12    & \citet{gaiadr2} Period &    days\\
PerCat13    & \citet{gaiadr2} Period &    days\\
PerCat14    & \citet{macho_eb} Period &    days\\
PerCat15    & \citet{macho_lpv} Period &    days\\
PerCat16    & \citet{macho_var} Period &    days\\
PerCat17    & \citet{eros_var} Period &    days\\
PerCat18    & \citet{eros2_lpv} Period &   days \\
PerCat19    & \citet{gaiadr3_2022} Period &   days \\
PerCat20    & \citet{gaiadr3_2022} Period &   days \\
PerCat21    & \citet{gaiadr3_2022} Period &   days \\
PerCat22    & \citet{vista_cep} Period & days \\
PerCat23    & \citet{vista_ecl} Period & days \\
\hline
\hline
\end{tabular}
\label{Table::knowncatnames}
\end{table}

A similar catalogue has been produced as well for the newly discovered variables, containing the Right Ascension and Declination columns, the period found with the Lomb-Scargle algorithm and associated error and power, the period found with the \czerny algorithm, the period found by the UPSILoN software with the associated uncertainty, the classification and the relative quality parameter both given by the UPSILoN analysis.
In Table \ref{Table::samplecatnewvar} we give an excerpt from the full catalogue. It must be noticed that the periods reported in this table may differ for each object, due to the different approach and technique used to obtain the estimate (for P1, P2 and P3). This difference arises from the computational limit of the algorithm or from the constraints adopted for the period search. For instance, the period P3 provided by the neural network is the one derived by the UPSILoN software itself to obtain the classification; hence, a different value between the latter and the other ones (P1 and P2) should not be surprising. As a consequence, this difference would lead to a classification in agreement with the P3 period but sometimes also discrepant with respect to the expectations from the P1 and P2 estimates.

Finally note that, in some particular cases the periodogram does not help in finding the period corresponding to a certain light curve, returning a default value close to 1.0 (often between about 0.99 and 1.01 days). For this reason, given the indefiniteness of these results, these periods have been indicated with NaN in Table \ref{Table::samplecatnewvar} for both the \czerny and the Lomb-Scargle periodograms.

\begin{table*}[h]
\caption{An excerpt of the first 20 objects, out of the whole set of 1266 new detected variables, is reported. In particular, columns from left to right report the coordinates RA\_ICRS and DE\_ICRS, P1 is the \czerny estimated period, P2 is the Lomb-Scargle estimated period with eP2 and powP2 the associated error and power of the major peak in the periodogram, P3 is the period estimate provided by the UPSILoN software and eP3 the associated error. All the estimated periods and the uncertainties are reported in days. The \textit{Class} column contains the classification for each variable and Prob indicates the estimated probability of this classification. The last column, \textit{CEPH\_flag} is equals to 1 if the Cepheid candidate satisfies the Period-Luminosity relation, otherwise the flag is always 0.}

\centering
\resizebox{\textwidth}{!}{\begin{tabular}{ccccccccccc}
\hline
\hline
\textbf{RA\_ICRS} & \textbf{DE\_ICRS} & \textbf{P1} & \textbf{P2} & \textbf{eP2} & \textbf{powP2} & \textbf{P3} & \textbf{eP3} & \textbf{Class}   & \textbf{Prob} & \textbf{CEPH\_flag} \\
deg &   deg &   days    &   days    &   days    &   & days & days &  &  & \\
\hline
\hline
80.21219          & -64.26791         & NAN         & NAN         & NAN          & NAN            & NAN         & NAN          & UNKNOWN          & NAN       &  0    \\
80.13829          & -64.17752         & 0.9885      & 0.1111      & 3e-06        & 0.3483         & 0.8653      & 0.000706     & EB\_ESD          & 0.37      &  0    \\
79.91440          & -64.25958         & 0.9743      & 0.9743      & 0.000161     & 0.5676         & 37.8859     & 1.66479      & EB\_ESD          & 0.33      &  0    \\
79.74037          & -64.17592         & 5.3272      & 0.1097      & 2e-06        & 0.4166         & 0.0836      & 7e-06        & NonVar           & 0.45      &  0    \\
79.90017          & -64.20354         & 0.8223      & 0.1088      & 2e-06        & 0.4282         & 0.098       & 9e-06        & NonVar           & 0.49      &  0    \\
81.51146          & -64.23720         & 1049.8126   & NAN         & NAN          & NAN            & 2121.6088   & 795.603313   & NonVar           & 0.34      &  0    \\
81.15678          & -64.21819         & 0.5937      & 0.5937      & 5e-05        & 0.6577         & 0.5938      & 0.000332     & RRL\_ab          & 0.4       &  0    \\
79.75284          & -64.36891         & 43.2121     & 19.7161     & 0.054711     & 0.5998         & 2121.6088   & 707.202945   & NonVar           & 0.55      &  0    \\
80.61756          & -64.37044         & 0.6084      & 0.5575      & 6.5e-05      & 0.4441         & 0.0336      & 1e-06        & DSCT             & 0.65      &  0    \\
81.79989          & -64.39814         & 66.0131     & 0.1105      & 2e-06        & 0.5959         & 38.5747     & 1.033922     & LPV\_AGB\_O      & 0.34      &  0    \\
78.80779          & -64.74936         & 0.6397      & 0.6397      & 7.5e-05      & 0.4823         & 0.0362      & 1e-06        & NonVar           & 0.47      &  0    \\
78.73691          & -64.77712         & 0.2015      & 0.1182      & 3e-06        & 0.4956         & 0.0567      & 3e-06        & DSCT             & 0.74      &  0    \\
79.82700          & -64.68342         & 876.699     & NAN         & NAN          & NAN            & 1060.8044   & 848.643534   & LPV\_AGB\_C      & 0.52      &  0    \\
79.39509          & -64.68179         & 19.706      & 19.7161     & 0.083769     & 0.4597         & 0.9493      & 0.000849     & CEPH\_Other      & 0.5       &  0    \\
79.38749          & -64.75647         & 0.3112      & 0.3112      & 1.7e-05      & 0.5887         & 0.0713      & 6e-06        & EB\_ESD          & 0.37      &  0    \\
80.56070          & -64.68838         & 1.3109      & 0.1006      & 2e-06        & 0.5247         & 0.0772      & 7e-06        & EB\_ESD          & 0.55      &  0    \\
80.41104          & -64.71999         & 0.2499      & 0.1999      & 5e-06        & 0.7057         & 0.1999      & 3.8e-05      & DSCT             & 0.64      &  0    \\
81.04415          & -64.73237         & 141.9728    & 145.2357    & 3.273094     & 0.6227         & 141.4406    & 19.20008     & LPV\_AGB\_C      & 0.36      &  0    \\
81.05842          & -64.78880         & 0.8744      & 0.8743      & 0.000123     & 0.5363         & 0.0358      & 1e-06        & DSCT             & 0.47      &  0    \\
81.04688          & -64.71910         & 0.2247      & 0.2247      & 9e-06        & 0.5699         & 0.0645      & 5e-06        & EB\_ESD          & 0.39      &  0    \\
\hline
\hline
\end{tabular}}
\label{Table::samplecatnewvar}
\end{table*}

\section{Conclusions}
\label{sec-conc}

In this paper we presented the analysis performed for the detection of variable sources in several  crowded fields (covering about 15 deg$^2$ of the sky) toward the LMC. The key result is the identification of 70~981 variable sources. Among these, 69 715 were already known since most of them have been previously detected  by different surveys such as OGLE, GAIA, EROS, MACHO and VISTA. This result clearly shows the capability of the adopted procedure to detect variable sources. 
For the known variable sources we cross-correlated our sources with those present in many catalogues available in literature. In particular, for each source we estimated the period and correlated it with the one available in the catalogues. 
It must be noticed, however, that the period of a fraction of the detected variable sources is not compatible with that available in the literature. In most cases this discrepancy is due to the sampling and the temporal limitation of the DECam data. In fact, our light curves have 20 -- 40 data points covering $\sim$2 years, while most of the other surveys, such as Gaia or OGLE, have a more dense sampling which allows an easier recognition of short period variables such as the RR-Lyrae, Cepheids or Eclipsing Binaries that populate the LMC.

In addition, considering the previously unknown variable sources, our procedure applied to the DECam data allowed the identification of 1266 new variables. As shown in Figure \ref{Fig:newvar-periods-corr} a good period-period correlation, between the \czerny and Lomb-Scargle period estimates, can be appreciated for a half of the new detected variables.
Additionally, an attempt of classification has been provided using the UPSILoN neural network which returns the estimated class and a quality parameter corresponding to the goodness of the provided classification. Most of these new variables, 413 among 1266, are recognized as Eclipsing Binaries (either contact, detached or semi-detached binaries), 98 are likely Cepheids - of which 53 agree with the Period-Luminosity relation - and 116 are possibly RR-Lyrae stars. Other variable sources are reported in Table \ref{Table::var_classes}. It must be noticed that, among the 1266 variables, 142 of these have been classified as {\it Non-Variable} and 43 have no classification ({\it UNKNOWN} in the catalogue) due to the aforementioned computational issues. For all these 185 sources we cannot claim a classification even if the \isis software would seem to indicate an almost certain variability, not always associated to a clear periodicity, as indicated by both the Lomb-Scargle and the \czerny periodograms.


\section*{Acknowledgements}

This paper is based on publicly available observations by DECam (Dark Energy Camera), an instrument mounted on the V. Blanco Telescope, as part of the Cerro Tololo Inter-American Observatory (Chile). 
We thank for partial support the INFN projects TAsP and EUCLID. 

\section*{Data Availability}

\noindent DECam images used for this work are publicly available at the https://astroarchive.noirlab.edu/portal/search/\#/search-form webpage.




\begin{thebibliography}{99}

\bibitem[Alcock et al. (1996)]{macho_micro}
Alcock, C. et al. 1996, AJ, 111, 1146.

\bibitem[Alcock et al. (2001)]{macho_var}
Alcock, C. et al. 2003, VizieR On-line Data Catalog: II/247.

\bibitem[Alard (2000)]{alard00} 
Alard, C. 2000, A\&AS, 144, 2, pp.363–370.

\bibitem[Alard \& Lupton (2000)]{alard98} 
Alard, C. \& Lupton, R. H. 1998, ApJ, 503, 1, pp. 325-331.

\bibitem[Derekas et al. (2007)]{macho_eb} 
Derekas, A. et al. 2007, ApJ, 663, 249.

\bibitem[Di Fabrizio et al. (2005)]{LMC_var}
Di Fabrizio, L. et al. 2005, A\&A 430, 603–628.

\bibitem[Flaugher et al. (2015)]{flaugher}
Flaugher, B. et al. 2015, AJ, 150, 150.

\bibitem[Frampton \& Chapline (2016)]{framp}
Frampton, P. H. \& Chapline, G. F. 2016, JCAP 11, 042.

\bibitem[Fraser et al. (2008)]{macho_lpv} 
Fraser, O. J. et al. 2008, AJ, 136, 1242.

\bibitem[Freedman et al. (2001)]{freedman2001} 
Freedman, W. L. et al. 2001, ApJ, 553, 47.

\bibitem[Gaia Collaboration (2018)]{gaiadr2} 
Gaia Collaboration: Brown A. G. A. et al 2018, A\&A 616, A1.

\bibitem[Gaia Collaboration (2020)]{gaiadr3}
Gaia Collaboration: Brown A. G. A., et al 2020, A\&A, 649, id.A1, 20 pp.

\bibitem[Gaia Collaboration (2022)]{gaiadr3_2022}
Gaia Collaboration: Brown A. G. A., et al 2022, A\&A, eprint arXiv:2208.00211.

\bibitem[Graczyk et al. (2011)]{ogle_eb} 
Graczyk, D. et al. 2011, Acta Astron., 61, 2, p. 103-122.

\bibitem[Kim et al. (2014)]{eros_var}
Kim, DW et al. 2014, A\&A, 566, A43.

\bibitem[Kim \& Bailer-Jones (2016)]{upsilon} 
Kim, DW \& Bailer-Jones, C. A. L. 2016 A\&A, 587, id.A18, 15 pp.

\bibitem[Lomb (1976)]{lomb} 
Lomb, N. R. 1976, Ap\&SS, pp. 447-462.

\bibitem[Muraveva et al. (2014)]{vista_ecl} 
Muraveva, T. et al. 2014, MNRAS, 443, 1, 432-445

\bibitem[Pawlak et al. (2016)]{ogle_eb2} 
Pawlak, M. et al. 2016, Acta Astron., 66, 4, p. 421-432.

\bibitem[Percy (2007)]{percy} 
Percy, J. 2007, Cambridge University Press, ISBN:9780521232531.

\bibitem[Pietrukowicz et al. (2017)]{ogle_blapss} 
Pietrukowicz, P. et al. 2017, Nature Astronomy, 1, 0166.

\bibitem[Poleski et al. (2010a)]{ogle_delta} 
Poleski, R. et al. 2010a, Acta Astron., 60, 1, p. 1-16.

\bibitem[Poleski et al. (2010b)]{ogle_double} 
Poleski, R. et al. 2010b, Acta Astron., 60, 3, p. 179-196.

\bibitem[Riello et al. (2021)]{riello}
Riello, M. et al. A\&AS, 2021, 649, id.A3, 33 pp.

\bibitem[Ripepi et al. (2012)]{vista_cep}
Ripepi, V. et al. MNRAS, 2012, 424, 1807-1816

\bibitem[Scargle (1982)]{scargle}
Scargle, J. D. 1982, ApJ, 1, 263, pp. 835-853.

\bibitem[Schwarzenberg-Czerny (1991)]{czerny1}
Schwarzenberg-Czerny, A. 1991, MNRAS, 253, 2, p.198-206.

\bibitem[Schwarzenberg-Czerny (1996)]{czerny2}
Schwarzenberg-Czerny, A. 1996, ApJ, 460, 2, p.L107-L110.

\bibitem[Soszy\'nski et al. (2008a)]{ogle_cep2}
Soszy\'nski, I. et al. 2008a, Acta Astron., 58, p. 293.

\bibitem[Soszy\'nski et al. (2008b)]{ogle_cep}
Soszy\'nski, I. et al. 2008b, Acta Astron., 58, pp. 163-185.

\bibitem[Soszy\'nski et al. (2009a)]{ogle_rrl}
Soszy\'nski, I. et al. 2009a, Acta Astron., 59, 1, p. 1-18.

\bibitem[Soszy\'nski et al. (2009b)]{ogle_rcb}
Soszy\'nski, I. et al. 2009b, Acta Astron., 59, 4, p. 335-347.

\bibitem[Soszy\'nski et al. (2011)]{ogle_lpv}
Soszy\'nski, I. et al. 2011, Acta Astron., 61, 3, p. 217-230.

\bibitem[Soszy\'nski (2018)]{ogle_onemill}
Soszy\'nski, I. 2018, Proc. of the Pol. Astr. Soc., vol. 7, 168-174.

\bibitem[Spano et al. (2011)]{eros2_lpv}
Spano, M. et al. 2011, A\&A 536, A60.

\bibitem[Stetson (1996)]{daophot}
Stetson, P. B. 1996, Users manual for DAOPHOT II.

\bibitem[Stringer et al. (2019)]{des_rrl}
Stringer, K. M. et al. 2019, AJ, 158, 16.

\bibitem[Udalski et al. (1997)]{ogle2}
Udalski, A. et al 1997, Acta Astron., 47, 319.

\bibitem[VanderPlas (2018)]{lombscargle2}
VanderPlas, J. T. 2018, APJS, 236, 1, id. 16.



\end{thebibliography}
\end{document}